\UseRawInputEncoding
\documentclass[pre,aps,amsmath,amssymb,amsfonts,floatfix,superscriptaddress,showpacs,twocolumn,footinbib]{revtex4-1}
\usepackage{graphicx}
\usepackage{amsmath,amsfonts}
\usepackage{xcolor}
\usepackage{color}
\usepackage{mathtools}
\usepackage{eufrak}
\usepackage{pgfplots}
\usepackage{soul}
\usepackage{multirow}
\newcommand{\up}{\uparrow}
\newcommand{\dn}{\downarrow}
\newcommand{\kv}{\ensuremath{\mathbf{k}}}
\newcommand{\lv}{\ensuremath{\boldsymbol{\ell}}}
\newcommand{\qv}{\ensuremath{\mathbf{q}}}

\newcommand{\ch}{\ensuremath{\text{ch}}}
\newcommand{\sz}{\ensuremath{\text{sp}}}

\newcommand{\pp}{\ensuremath{{pp}}}

\newcommand{\trip}{\ensuremath{\text{t}}}
\newcommand{\sing}{\ensuremath{\text{s}}}
\newcommand{\firr}{\ensuremath{\text{Uirr}}}

\usepackage{tikz}
\usetikzlibrary{calc}
\usetikzlibrary{decorations.pathmorphing}
\usetikzlibrary{decorations.pathreplacing}
\usetikzlibrary{decorations.markings}
\usetikzlibrary{shapes.misc}
\usetikzlibrary{shapes}
\usetikzlibrary{positioning}
\usetikzlibrary{snakes}
\usetikzlibrary{arrows}
\usetikzlibrary{fadings}
\usetikzlibrary{calc,arrows}
\tikzstyle{decision} = [diamond, draw, fill=blue!20, text width=4.5em, text badly centered, node distance=3cm, inner sep=0pt]
\tikzstyle{block} = [rectangle, draw, fill=blue!20, text width=7em, text centered, rounded corners, minimum height=5em]
\tikzstyle{line} = [draw, -latex']
\tikzstyle{cloud} = [draw, ellipse,fill=red!20, node distance=3cm, minimum height=2em]
\tikzstyle{overbrace style}=[decorate,decoration={brace,raise=2mm,amplitude=3pt}]
\tikzstyle{overbrace text style}=[font=\footnotesize, above, pos=.5, yshift=3mm]
\tikzset{snake it/.style={decorate, decoration=snake}}
\usetikzlibrary{3d}
\usepackage{mathtools}
    \tikzset{
            partial ellipse/.style args={#1:#2:#3}{
                        insert path={+ (#1:#3) arc (#1:#2:#3)}
                            }
                        }

\usetikzlibrary{calc}
\tikzset{
            inertial frame/.style = {x={(-20:2cm)}, y={(-160:2cm)}, z={(90:2cm)}},
              local frame/.style = {shift={(local origin)}, x={(40:.7cm)}, y={(150:.7cm)}, z={(105:.7cm)}}
          }

    \tikzset{middlearrow/.style={
                decoration={markings,
                            mark= at position 0.65 with {\arrow{#1}} ,
                                    },
                                            postaction={decorate}
                                                }
                                                }
\tikzset{cross/.style={cross out, draw, 
         minimum size=2*(#1-\pgflinewidth), 
                  inner sep=0pt, outer sep=0pt}}

\def\presuper#1#2%
  {\mathop{}%
   \mathopen{\vphantom{#2}}^{#1}%
   \kern-0.5\scriptspace%
   #2}

\usepackage{hyperref}
\hypersetup{colorlinks=true,breaklinks,linkcolor=blue,urlcolor=blue,citecolor=blue}
\begin{document}
%\pgfplotsset{compat=1.8}
%\pgfmathdeclarefunction{gauss}{3}{%
%    \pgfmathparse{1/(#3*sqrt(2*pi))*exp(-((#1-#2)^2)/(2*#3^2))}%
%    }

    \pgfmathdeclarefunction{gauss}{2}{%
          \pgfmathparse{1/(#2*sqrt(2*pi))*exp(-((x-#1)^2)/(2*#2^2))}%
          }
    \pgfmathdeclarefunction{mgauss}{2}{%
          \pgfmathparse{-1/(#2*sqrt(2*pi))*exp(-((x-#1)^2)/(2*#2^2))}%
          }
    \pgfmathdeclarefunction{lorentzian}{2}{%
        \pgfmathparse{1/(#2*pi)*((#2)^2)/((x-#1)^2+(#2)^2)}%
          }
    \pgfmathdeclarefunction{mlorentzian}{2}{%
        \pgfmathparse{-1/(#2*pi)*((#2)^2)/((x-#1)^2+(#2)^2)}%
          }

\author{Friedrich Krien}
%\email{friedrich.krien@ijs.si}
\email{krien@ifp.tuwien.ac.at}
\affiliation{Jo\v{z}ef Stefan Institute, Jamova 39, SI-1000, Ljubljana, Slovenia}
\affiliation{Institute for Solid State Physics, TU Wien, 1040 Vienna, Austria}

\author{Angelo Valli}
\affiliation{Institute for Theoretical Physics, TU Wien, 1040 Vienna, Austria}

\author{Patrick Chalupa}
\affiliation{Institute for Solid State Physics, TU Wien, 1040 Vienna, Austria}

\author{Massimo Capone}
\affiliation{International School for Advanced Studies (SISSA), Via Bonomea 265, 34136 Trieste, Italy}
\affiliation{CNR-IOM Democritos, Via Bonomea 265, 34136 Trieste, Italy}

\author{Alexander I. Lichtenstein}
\affiliation{Institute of Theoretical Physics, University of Hamburg, 20355 Hamburg, Germany}

\author{Alessandro Toschi}
\affiliation{Institute for Solid State Physics, TU Wien, 1040 Vienna, Austria}

%\author{}
%\affiliation{}

\title{Boson Exchange Parquet Solver for dual fermions}

\begin{abstract}
{We present and implement a parquet approximation within the dual-fermion formalism based
on a partial bosonization of the dual vertex function which substantially
reduces the computational cost of the calculation.
The method relies on splitting the vertex exactly into single-boson exchange
contributions and a residual four-fermion vertex,
which physically embody respectively long-range and short-range spatial correlations.}
After recasting the parquet equations in terms of the residual vertex, these are solved
using the truncated unity method of Eckhardt et al.
[\href{https://journals.aps.org/prb/abstract/10.1103/PhysRevB.101.155104}{Phys. Rev. B 101, 155104 (2020)}],
which allows for a rapid convergence with the number of form factors in different regimes.
While our numerical treatment of the parquet equations can be restricted to only a few Matsubara frequencies, reminiscent of Astretsov et al. [\href{https://journals.aps.org/prb/abstract/10.1103/PhysRevB.101.075109}{Phys. Rev. B 101, 075109 (2020)}], the one- and two-particle spectral information is fully retained.
In applications to the two-dimensional Hubbard model
the method agrees quantitatively with a stochastic summation of diagrams over a wide range of parameters.
\end{abstract}

\maketitle

\section{Introduction}
The two-dimensional Hubbard model even with a single band still poses a formidable challenge to theorists.
Despite an immense collective effort, which led to the development of many novel methods,
the model has not been solved exactly and no approximate method works accurately in every regime.
Arguably, one of the most delicate, and at the same time most interesting,
parameter regime is realized in the doped Hubbard model
at low-temperatures and for intermediate-to-strong coupling interactions, which is precisely 
the regime of relevance for the low-energy modelization of the cuprate~\cite{Andersen1995} and,
as recently suggested, of the nickelate superconductors~\cite{Li19,Kitatani20}.

In this region there is no natural small parameter and perturbative approaches are bound to fail.
Most of the features of the cuprate phase diagram, like the pseudogap behavior of spectral~\cite{Timusk1999,Wu18} and transport properties~\cite{Badoux2016,Michon2018}, $d$-wave superconductivity~\cite{Scalapino12,Otsuki14,Kitatani19},
and a plethora of other exotic phenomena such as unconventional density waves~\cite{Webb19},
stripe order~\cite{Zheng17,Jiang19}, phase separation~\cite{Otsuki14,Nourafkan18,Reitner20},
or a $T$-linear resistivity~\cite{Huang18,Brown19} have been reported.

The impossibility to apply conventional small-parameter expansion schemes, makes it necessary to resort to non-perturbative approaches. 
In this regard  dynamical mean-field theory (DMFT)~\cite{Georges96}, which approximates the self-energy with a local version which can be computed
from a self-consistent impurity model, has become a reference method.
Standard DMFT can not capture the momentum-dependent physics of two-dimensional systems, calling for cluster extensions, like the dynamical cluster approximation (DCA)~\cite{Hettler98} or the cellular-DMFT~\cite{Lichtenstein00,Kotliar01}.
However, some relevant aspects of the two-dimensional physics can not be captured by cluster methods, which are limited to fairly small clusters and therefore include only short-ranged correlations.
For instance, we refer here to the description of unconventional charge-density, spin-density waves and pseudogap features~\cite{Timusk1999,Badoux2016}, or, more in general, to the treatment of long-range spatial correlations -- a typical hallmark~\cite{Vilk97,Schaefer15,Rohringer16,Rohringer18,Schaefer2020}
of strongly correlated physics in two-dimensions. A proper treatment of these phenomena
intrinsically requires a fine resolution of the Brillouin zone, which could be obtained in cluster DMFT only for very large clusters beyond any practical implementation.

Diagrammatic extensions of DMFT~\cite{Rohringer18} aim at  including spatial correlations beyond DMFT. Here it is important to make the methods as cheap as possible from a computational point of view,  
so  that the number of lattice momenta can be kept large.

In this framework, approaches based on the ladder approximation~\cite{Toschi07,Hafermann09}
allow for a high-resolution in momentum space for the half-filled Hubbard model.
In this regime it is known \textit{a priori} that spin fluctuations are dominant,
explaining the accuracy of the corresponding ladder-treatments.
Away from half-filling the situation becomes more complex, as the physics turns out to be
controlled by a delicate interplay between bosonic fluctuations in different channels ~\cite{Pudleiner19-2} even if
spin fluctuations still play a pivotal role in determining single-particle spectral properties~\cite{Gunnarsson15},
possibly with significant renormalization effects arising from other scattering channels~\cite{Gunnarsson16}.

A very general way to describe this interplay is to take parquet-type diagrams into 
account~\cite{Dominicis64,Dominicis64-2,Bickers04,Yang09,Tam13,Rohringer12,Valli15,Schueler17,Pudleiner19,Kauch19,Kauch19-2}. However, due to the heavy numerical cost of the parquet equations, it appeared so far impossible to achieve a spatial resolution comparable to that of the ladder approximations.

Recent papers have reported improvements in this direction. 
First, Astretsov et al.~\cite{Astretsov19} combined the dual fermion (DF) approach~\cite{Rubtsov08}
with the renormalization group (RG),
treating parquet diagrams only for the two smallest Matsubara frequencies (we refer to this as DP+RG).
As a result, one can work with large clusters, up to $32\times32$ sites in the mentioned manuscript.
Second, Eckhardt et al.~\cite{Eckhardt20} applied the truncated-unity~\cite{Lichtenstein17}
form-factor expansion to the parquet equations (TUPS),
which corresponds to a truncated real-space representation of the vertex function $F(k,k',q)$ with respect to its two fermionic momentum arguments $\kv$ and $\kv'$, where $k=(\kv,\nu)$ denotes a momentum-energy.
This approximation corresponds to the assumption of a short-ranged dependence of the vertex $F$ on the fermionic momenta. When this condition is satisfied, the truncated unity allows for a very large lattice size and retains the full spectral information encoded in the Matsubara frequencies.

In this work, we contribute to the current progress by addressing two specific problems which arise in the DP+RG and TUPS methods. On the one hand, the RG treatment neglects spectral information from higher Matsubara frequencies and, hence,
it is not straightforward to obtain the spectral density (DOS) or susceptibilities.
On the other hand, the convergence of observables in TUPS with the number of form factors can be slow in the regime of strong spatial correlations.

\begin{figure}
\begin{center}
\includegraphics[width=0.45\textwidth]{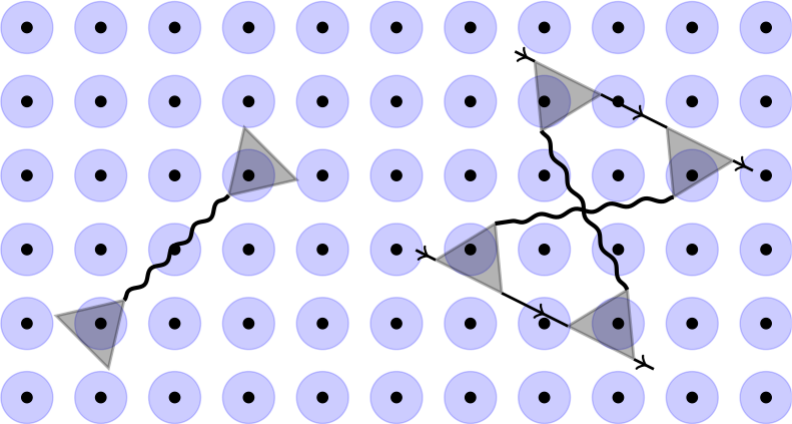}
\end{center}
    \caption{\label{fig:lattice}
    Schematic representation of the BEPS method.
    The Hubbard model is mapped to a collection of impurities
    embedded in a self-consistent bath (blue circles), which account for local correlations.
    Nonlocal correlations are added in a dual perturbation theory.
    Interaction between dual fermions (arrows) is mediated by
    bosons (wiggly lines) and a fermion-boson coupling (triangles).
    Left: Maki-Thompson correction. Right: Aslamazov-Larkin correction.
    }
    \end{figure}

In this work we propose a scheme based on the parquet
approximation for dual fermions~\cite{Rubtsov08,Astretsov19,Astleithner20}
which overcomes the limitations of the two mentioned approaches. The method exploits
a partial bosonization~\cite{Krahl07,Friederich10,Bartosch09,Streib13,Denz19}
of the dual vertex function,
similar to the channel decomposition~\cite{Karrasch08,Husemann09,Honerkamp18,Vilardi19}
used in the context of the functional renormalization group (fRG,~\cite{Metzner12})
or in the microscopic Fermi liquid theory~\cite{Reidy14}.

The partial bosonization is performed in terms of the recently introduced
exact single-boson exchange (SBE) decomposition of the vertex function~\cite{Krien19-2}.
The single-boson exchange corresponds to fluctuations which couple to the bare interaction of the Hubbard model, and they completely characterize the vertex at high frequencies~\cite{Wentzell2020}.
If we write the full vertex as the sum of the single-boson exchange terms
and of an irreducible term $\Phi^\firr$, the latter 
is a residual four-fermion vertex whose frequency and momentum structure is simplified in two important ways.
First, $\Phi^\firr$ decays for high energies in {\sl all} directions of the Matsubara frequency space~\cite{Krien19-2}.
This is somewhat similar to the asymptotic behavior of the fully 2PI vertex in standard parquet approaches~\cite{Rohringer12,Wentzell2020}. At the same time, $\Phi^\firr$ appears {\sl not} to be affected by the multiple strong-coupling divergences~\cite{Schaefer13,Gunnarsson16,Schaefer16,Gunnarsson17,Chalupa18,Thunstroem18,Springer20,Chalupa20}
which otherwise make the numerical treatment of 2PI vertices problematic.
{In this respect, we note that one of the advantages of implementing  parquet-based approximations in the dual-fermion formalism is the possibility of avoiding, at any stage of the procedure, to manipulate local 2PI vertex functions~\footnote{For the 2PI vertices in selected channels, this was already discussed in Refs.~\cite{Rohringerthesis,Rohringer18,vanLoon20}}, while fully retaining the whole {\sl non-perturbative} information that they encode~\cite{Chalupa20,Reitner20}.}

{Further, it should be also emphasized that, in general, $\Phi^\firr$ is found to be} significantly shorter-ranged in space compared to the full vertex function,
because many typical long-ranged correlations (such as spin- and charge-density wave),
correspond to single-boson exchange.

In this work we exploit these properties by formulating a truncated unity parquet solver similar to Ref.~\cite{Eckhardt20}
for the residual four-fermion vertex $\Phi^\firr$.
Since this vertex describes low-energy and short-ranged correlations
we achieve a fast convergence of the parquet diagrams with respect to Matsubara sums 
and in terms of the form-factor expansion,
making the {\sl converged} solution of the parquet equations
much less computationally demanding compared to previous calculation schemes.

Our exact reformulation of the dual parquet equations requires the introduction of
bosonic lines, which are given by the screened interaction, and a fermion-boson coupling
(the {\sl Hedin vertex}~\cite{Hedin65}, see, e.g.,~\cite{Schmalian99,Abanov03,Katanin09,Sadovskii19,Krien19}).
In the dual fermion formalism the lattice quantities can be expressed as the sum of local
and nonlocal contributions~\cite{Rubtsov08}. In this spirit,
we express the fermion-boson coupling as the sum of the local impurity quantity plus corrections,
\begin{align}
  \Lambda(k,q)=\lambda^\text{imp}(\nu,\omega)+\Lambda^\text{nonloc}(k,q).
\end{align}
Since the local approximation $\Lambda\approx\lambda$ corresponds to the TRILEX approach~\cite{Ayral15},
our method can also be seen as a crossing-symmetric extension of TRILEX.
An exact prescription for the renormalization of the fermion-boson coupling was recently presented in Ref.~\cite{Krien19-3} for lattice fermions. In this work we extend this concept to dual variables and show
how the parquet diagrams can be systematically expressed in terms of boson exchange,
such as the Maki-Thompson (single-boson exchange)
and the Aslamazov-Larkin (two-boson exchange) vertex corrections, shown in Fig.~\ref{fig:lattice}.
They illustrate our targeted application of the truncated unity:
The strong momentum-dependence of the Maki-Thompson diagram is fully retained,
whereas the more short-ranged spatial dependence of the Aslamazov-Larkin diagram
is captured by a small number of form factors, making the method computationally feasible.
In the following, we refer to this method as a \textit{boson exchange parquet solver} (BEPS).

The paper is structured as follows. We introduce the BEPS method in Sec.~\ref{sec:method}.
We benchmark the method at half-filling against diagrammatic Monte Carlo in Sec.~\ref{sec:benchmark},
we discuss the doped case in Sec.~\ref{sec:benchmark_doped}. We conclude in Sec.~\ref{sec:conclusions}.

\section{Model and Method}\label{sec:method}
\subsection{Hubbard model}
In the applications we consider the paramagnetic Hubbard model on the square lattice,
\begin{align}
    H = &-t\sum_{\langle ij\rangle\sigma}c^\dagger_{i\sigma}c^{}_{j\sigma}+ U\sum_{i} n_{i\up} n_{i\dn},\label{eq:hubbardmodel}
\end{align}
where $t=1$ is the nearest neighbor hopping which sets the unit of energy.
$c^{},c^\dagger$ are the annihilation and creation operators, $\sigma=\up,\dn$ the spin index.
$U$ is the Hubbard repulsion between the densities $n_{\sigma}=c^\dagger_{\sigma}c^{}_{\sigma}$.
The spin label $\sigma$ is suppressed where unambiguous.

\subsection{Anderson impurity model}\label{sec:aim}
Our method is based on an auxiliary Anderson Impurity Model (AIM) with the imaginary time action,
\begin{align}
  S_{\text{AIM}}=&-\sum_{\nu\sigma}c^*_{\nu\sigma}(\imath\nu+\mu-h_\nu)c^{}_{\nu\sigma}+U \sum_\omega n_{\up\omega} n_{\dn\omega},
  \label{eq:aim}
\end{align}
where $c^*,c$ are Grassmann numbers, $\nu$ and $\omega$ 
are fermionic and bosonic Matsubara frequencies, respectively.
Summations over Matsubara frequencies $\nu, \omega$ contain implicitly the factor $T=\beta^{-1}$,
the temperature. In our scheme, the auxiliary AIM is exploited to solve the lattice problem
under investigation within the dynamical mean-field theory (DMFT), which represents 
the starting point of our analysis.

The specific hybridization function $h_\nu$ of our AIM corresponds, thus, to the self-consistent DMFT solution~\cite{Georges96} for the Hubbard model~\eqref{eq:hubbardmodel} where the local part of the lattice Green's function is adjusted to the local Green's function $g_\sigma(\nu)=-\langle c^{}_{\nu\sigma}c^*_{\nu\sigma}\rangle$ of the AIM, $G^\text{DMFT}_{ii}(\nu)=g(\nu)$.

We require several higher correlation functions of the AIM~\eqref{eq:aim},
namely, the four-point function,
\begin{align}
    g^{(4),\alpha}_{\nu\nu'\omega}=&-\frac{1}{2}\sum_{\sigma_i}s^\alpha_{\sigma_1'\sigma_1^{}}s^\alpha_{\sigma_2'\sigma_2^{}}
    \langle{c^{}_{\nu\sigma_1}c^{*}_{\nu+\omega,\sigma_1'}c^{}_{\nu'+\omega,\sigma_2}c^{*}_{\nu'\sigma_2'}}\rangle\notag,
\end{align}
where $s^\alpha$ are the Pauli matrices and the label $\alpha=\ch,\sz$ denotes the charge and spin channel, respectively.
This defines the four-point vertex function $f$ as,
\begin{align}
f^\alpha_{\nu\nu'\omega}=&\frac{g^{(4),\alpha}_{\nu\nu'\omega}-\beta g_\nu g_{\nu+\omega}\delta_{\nu\nu'}
    +2\beta g_\nu g_{\nu'}\delta_{\omega}\delta_{\alpha,\ch}}{g_\nu g_{\nu+\omega}g_{\nu'}g_{\nu'+\omega}}\label{eq:4pvertex}.
\end{align}
Charge, spin, and singlet susceptibilities are given as,
\begin{align}
    \chi^\alpha_\omega=&-\langle{\rho^\alpha_{-\omega}\rho^\alpha_\omega}\rangle+\beta\langle n\rangle\langle n\rangle\delta_\omega\delta_{\alpha,\ch},\\
    \chi^\sing_{\omega}=&-\left\langle \rho^-_{-\omega}\rho^{+}_{\omega}\right\rangle,\label{eq:chi_sing}
\end{align}
where $\rho^\ch=n_\up+n_\dn=n$ and $\rho^\sz=n_\up-n_\dn$ in the first line are the charge and spin densities 
whereas $\rho^+=c^{*}_\up c^{*}_\dn$ and $\rho^-=c_\dn c_\up$ describe the creation and annihilation of an electron pair. From the susceptibility we obtain the screened interaction,
\begin{eqnarray}
    w^\alpha_\omega=U^\alpha+\frac{1}{2}U^\alpha\chi^\alpha_\omega U^\alpha,\label{eq:w}
\end{eqnarray}
where $U^\ch=U, U^\sz=-U, U^\sing=2U$ is the bare interaction of the Hubbard model~\eqref{eq:hubbardmodel}
in the respective channel. Finally, we define the fermion-boson coupling of the impurity as~\footnote{
The \textit{reducible} vertex $\lambda^\text{red}$,
without $w^\alpha(\omega)/U^\alpha$ in the denominator, is discussed in Ref.~\cite{vanLoon18}.},
\begin{align}
    {\lambda}^{\alpha}_{\nu\omega}=\frac{
        \frac{1}{2}\sum_{\sigma\sigma'}s^\alpha_{\sigma'\sigma}\langle{c^{}_{\nu\sigma}c^{*}_{\nu+\omega,\sigma'}\rho^\alpha_\omega}\rangle
        +\beta g_\nu \langle n\rangle\delta_{\omega}\delta_{\alpha,\ch}}
    {g_\nu g_{\nu+\omega}w^\alpha_\omega/U^\alpha},\label{eq:hedinvertex}
\end{align}
for the particle-hole channels, $\alpha=\ch, \sz$, and
\begin{align}
    {\lambda}^{\sing}_{\nu\omega}=\frac{
    \left\langle c_{\nu\up}c_{{\omega}-\nu,\dn}\rho^{+}_{{\omega}}\right\rangle
}{g_\nu g_{\omega-\nu}w^\sing_{\omega}/U^\sing},\label{eq:lambdasing}
\end{align}
for the singlet particle-particle channel, $\alpha=\sing$.

In the single-boson exchange (SBE) decomposition~\cite{Krien19-2} the full vertex $f$
is split into three vertices $\nabla$ which are reducible with respect to the bare interaction $U$,
and one residual four-fermion vertex $\varphi^\firr$, irreducible w.r.t.~$U$,
\begin{align}
    f^\alpha_{\nu\nu'\omega}\!=\!\varphi^{\firr,\alpha}_{\nu\nu'\omega}\!+\!\nabla^{ph,\alpha}_{\nu\nu'\omega}
    \!+\!\nabla^{\overline{ph},\alpha}_{\nu\nu'\omega}\!+\!\nabla^{\pp,\alpha}_{\nu\nu',\omega+\nu+\nu'}\!-\!2U^\alpha\!.\label{eq:jib_aim}
\end{align}
Note that the bare interaction $U^\alpha$ is subtracted twice as a double counting correction,
which leads to the correct high-frequency asymptotics of $f$.

The ($U$-reducible) vertices $\nabla$ are given by the \textit{screened} interaction
$w$ and the fermion-boson coupling $\lambda$,
\begin{align}
\nabla^\alpha_{\nu\nu'\omega}=\lambda^\alpha_{\nu\omega}w^\alpha_\omega\lambda^\alpha_{\nu'\omega},\label{eq:nabla_aim}
\end{align}
where $\alpha=\ch,\sz,\sing$. We discuss their meaning in more detail in Sec.~\ref{sec:strategy}
for the lattice Hubbard model~\eqref{eq:hubbardmodel}.

The $U$-irreducible vertex $\varphi^\firr$ represents, instead, a natural starting point for approximations~\cite{Krien19-3} of more complex many-electron problems on a lattice,
as it is also the case in this work.
We obtain it through Eq.~(\ref{eq:jib_aim}), after measuring the 
vertices in Eqs.~\eqref{eq:4pvertex},~\eqref{eq:hedinvertex}, and~\eqref{eq:lambdasing}
with a continuous-time quantum Monte-Carlo (CTQMC) solver~\cite{Gull11,ALPS2,Wallerberger19}
with improved estimators~\cite{Hafermann12}.
These pieces are used to form the vertices $\nabla$ which are subtracted from
the full vertex $f$ to obtain $\varphi^\firr$.

\subsection{Dual fermions}
\begin{figure}
\begin{center}
\includegraphics[width=0.45\textwidth]{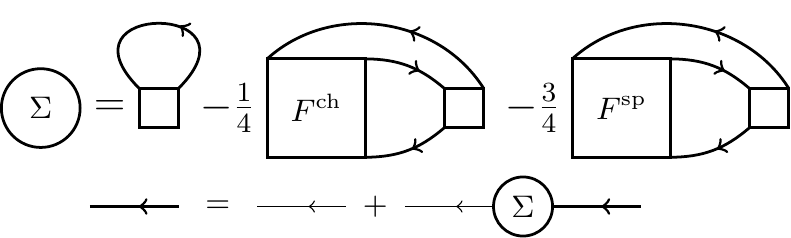}
\end{center}
    \caption{\label{fig:sigma}
    Top: Dual self-energy. Arrows denote the dual Green's function $G$,
    large boxes represent the vertex function $F$, small boxes the impurity vertex $f$.
    Bottom: Dyson equation, thin arrows represent the bare dual Green's function $G^0$.
    }
    \end{figure}

In the dual fermion formalism~\cite{Rubtsov08} the Hubbard model~\eqref{eq:hubbardmodel} is mapped to the dual action~\footnote{
We use a different sign convention for the vertex function $f$ than, e.g., Ref.~\cite{Otsuki14}.
As a result, the vertex is given to first order as $f^{\ch/\sz}=U^{\ch/\sz}+\mathcal{O}(U^2)=\pm U+\mathcal{O}(U^2)$},
\begin{align}
S[d^*,d]=&-\sum_{k\sigma}G^{0,-1}_kd^*_{k\sigma}d_{k\sigma}\label{eq:dualaction}\\
+&\frac{1}{4}\sum_{kk'q}\sum_{\sigma_i}f^{\sigma_1\sigma_2\sigma_3\sigma_4}_{\nu\nu'\omega}
d^*_{k\sigma_1}d^*_{k'+q,\sigma_2}d_{k'\sigma_3}d_{k+q,\sigma_4}.\notag
\end{align}
The Grassmann numbers $d^*,d$ represent the dual fermions and the
bare propagator is the nonlocal DMFT Green's function, $G^0=G^\text{DMFT}-g$.
A common approximation is to neglect higher than quartic interactions between the dual fermions,
the interaction is then given by the vertex $f$ of the AIM defined in Eq.~\eqref{eq:4pvertex}.
The bare propagator $G^0$ is then dressed with a dual self-energy,
\begin{align}
G_k=\frac{G^0_k}{1-G^0_k\Sigma_k}.\label{eq:dyson}
\end{align}
The self-energy reads in the general case (\cite{Hirschmeierthesis}, cf. Fig.~\ref{fig:sigma}),
\begin{align}
{\Sigma}_k=&\sum_{k'}{G}_{k'}f^\ch_{\nu'\nu,\omega=0}\label{eq:sigma}\\
-&\frac{1}{4}\sum_{k'q}{G}_{k+q}\left[{F}^\ch_{kk'q}{X}^0_{k'q}f^\ch_{\nu'\nu\omega}
+3{F}^\sz_{kk'q}{X}^0_{k'q}f^\sz_{\nu'\nu\omega}\right].\notag
\end{align}
Here, $X^0_{kq}=G_kG_{k+q}$ denotes a bubble of dual Green's functions and $F$ is the full vertex function of the dual fermions. It has the leading term $f$, the impurity vertex,
higher terms are all one-particle irreducible diagrams built from $f$ and the dual Green's function $G$~\cite{Astretsov19}.

After a self-consistent solution for $\Sigma_k$ is obtained,
we recover the approximation for the self-energy of the Hubbard model~\eqref{eq:hubbardmodel} as,
\begin{align}
\Sigma^\text{lat}_k=\Sigma^\text{DMFT}_\nu+\frac{\Sigma_k}{1+g_\nu\Sigma_k},\label{eq:sigma_lat}
\end{align}
where $\Sigma^\text{DMFT}_\nu$ and $g_\nu$ denote, respectively, the self-energy and local Green's function of the self-consistent DMFT solution of the Hubbard
model~\eqref{eq:hubbardmodel}, obtained from the corresponding auxiliary AIM~\eqref{eq:aim}.

\subsection{Strategy overview}\label{sec:strategy}
In the following we develop an efficient method for the summation of parquet diagrams.
We begin to explain our strategy by noting that recently an exact diagrammatic decomposition
was presented in Ref.~\cite{Krien19-2}, which separates diagrams from the vertex function that correspond to single-boson exchange. For the vertex function of the dual fermions this decomposition reads ($\alpha=\ch,\sz$),
\begin{align}
    F^\alpha_{kk'q}\!=\!\Phi^{\firr,\alpha}_{kk'q}\!+\!\Delta^{ph,\alpha}_{kk'q}
    \!+\!\Delta^{\overline{ph},\alpha}_{kk'q}\!+\!\Delta^{\pp,\alpha}_{kk',q+k+k'}
    \!-\!2U^\alpha.\!\!\label{eq:jib_simple}
\end{align}

Here, the vertices $\Delta$ represent the single-boson exchange of the dual fermions and
$\Phi^\firr$ denotes a four-fermion `rest' vertex,
analogous to the impurity quantities $\nabla$ and $\varphi^\firr$ previously introduced in Eq.~\eqref{eq:jib_aim}, respectively.
Hereafter, we will adopt in general capital (small) letters for lattice (impurity) quantities.
The label `Uirr' indicates that $\Phi^\firr$ does not have insertions of the bare interaction $U$~\cite{Krien19-2}.
The decomposition shares a similarity with the traditional parquet decomposition~\cite{Rohringer12,Gunnarsson16,Krien19-3} because single-boson exchange occurs in the horizontal ($ph$) and vertical ($\overline{ph}$) particle-hole channels and in the (singlet) particle-particle ($\pp$) channel. The SBE vertices have the structure shown in Fig.~\ref{fig:sbe},
\begin{subequations}
\begin{align}
    \Delta^{ph,\alpha}(k,k',q)=&\Lambda^\alpha(k,q)W^\alpha(q)\Lambda^\alpha(k',q),\label{eq:nabla_simple}\\
    \Delta^{pp,\sing\,}(k,k',q)=&\Lambda^{\sing\,}(k,q)W^{\sing\,}(q)\,\Lambda^{\sing\,}(k',q),\label{eq:nabla_simple_pp}
\end{align}
\end{subequations}
where $W$ denotes the screened interaction of the dual fermions and $\Lambda$ is the (dual) fermion-boson coupling,
see also Appendix~\ref{app:susc}.
We explain how the SBE decomposition~\eqref{eq:jib_simple} can help
to overcome two open problems that arise in the DP+RG and TUPS methods~\cite{Astretsov19,Eckhardt20}:

\begin{figure}[b]
\begin{center}
\includegraphics[width=0.25\textwidth]{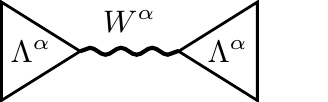}
\end{center}
\caption{\label{fig:sbe}
A vertex correction corresponding to single-boson exchange.
Triangles represent the fermion-boson coupling, the wiggly line denotes the screened interaction.}
\end{figure}
    
(i) The SBE vertices $\Delta$ control the asymptotics of the full vertex $F$~\cite{Wentzell2020,Kaufmann17,Krien19-3},
hence, the decomposition~\eqref{eq:jib_simple} helps to separate high from low energy scales. 
Consistent with this observation, in this work, we formulate the parquet equations for the four-fermion vertex $\Phi^\firr$ of the SBE decomposition~\eqref{eq:jib_simple}, restricting ourselves to a handful of Matsubara frequencies,
in the same spirit as the DP+RG ansatz of Astretsov et al.~\cite{Astretsov19}.
However, since $\Phi^\firr$ decays with respect to {\sl all} of its frequency arguments,
this can be done without a significant loss of spectral information,
whereas the DP+RG method omits information from Matsubara frequencies $|\nu|>\pi T$.

(ii) The boson $W(q)$ encodes the physics of long-ranged fluctuations, for example, the spin fluctuations
of the Hubbard model near half-filling~\cite{Schaefer15,vanLoon18-2}.
This explains the possible emergence of strong dependencies on the bosonic momentum $\qv$ in the full vertex $F$.

A procedure often used to simplify the treatment of the momentum dependence of two-particle diagrams~\cite{Husemann09,Wang12,Lichtenstein17}
is to expand the full vertex in terms of form factors,
\begin{align}
  F^\alpha(\ell,\ell',q)=\sum_{\kv\kv'}\psi(\lv,\kv) F^\alpha(k,k',q)\psi(\lv',\kv'),\label{eq:ff}
\end{align}
where $\psi$ denotes a form factor and $\ell=(\lv,\nu)$ is an appropriate multi-index
denoting form-factor index and Matsubara frequency.
Eq.~\eqref{eq:ff} is exact, but in the truncated unity approach only a few form factors are taken into account which capture short-ranged real space correlations~\cite{Thomale13,Eckhardt20}.
Typically, one uses a specific number of form factors, $N_\ell=1,5,9,13,...$,
which corresponds to truncation of the real space expansion after the zeroth
(1, corresponding to the local approximation),
first (5), second (9), third (13) nearest neighbors and so forth.

We note here that the truncation does not affect the momentum $\qv$ and is therefore appropriate
for the SBE vertex $\Delta^{ph}$ of the horizontal particle-hole channel in Eq.~\eqref{eq:nabla_simple}.
However, due to the crossing-symmetry, bosonic fluctuations contribute to $F$ also
in the vertical particle-hole channel ($\alpha=\ch,\sz$),
\begin{align}
 \Delta^{\overline{ph},\alpha}(k,k',q)=-\frac{1}{2}&\Delta^{ph,\ch}(k,k+q,k'-k)\label{eq:cs}\\
 -\frac{3-4\delta_{\alpha,\sz}}{2}&\Delta^{ph,\sz}(k,k+q,k'-k),\notag
\end{align}
and a further boson arises from singlet fluctuations,
\begin{align}
\Delta^{pp,\alpha}(k,k',q)=&\frac{1-2\delta_{\alpha,\sz}}{2}\Delta^{pp,s}(k,k',q).
\label{eq:nabla_singlet}
\end{align}

Equations~\eqref{eq:cs} and~\eqref{eq:nabla_singlet} indicate that a problem can arise from
a straightforward application of the truncated unity approximation to the full vertex $F$,
because it implies a (truncated-unity) cutoff also for bosonic fluctuations with momenta $\kv'-\kv$
and $\qv+\kv'+\kv$ [cf. Eq.~\eqref{eq:jib_simple}],
which may be long-ranged. 
Therefore, in our scheme, we exploit the truncated unity approximation only for the vertex
$\Phi^\firr$ in Eq.~\eqref{eq:jib_simple}, retaining the full momentum-dependence of the SBE vertices $\Delta$.
Indeed, the momentum-dependence of $\Phi^\firr$ is short-ranged,
leading to a faster convergence of the form-factor expansion, that is,
\begin{align}
\Phi^\firr({\ell},{\ell'},q)\approx0,
\end{align}
when $\lv$ or $\lv'$ correspond to long distances in the real space.

\subsection{Parquet expressions for the residual vertex}\label{sec:p3}
In Ref.~\cite{Eckhardt20} the TUPS was introduced to reduce the algorithmic
complexity of the parquet equations for the full vertex function $F$.
Here, {our aim is to} apply the TUPS to the residual vertex $\Phi^\firr$ {only}.
{Hence,} as {anticipated} in the previous section, 
we need to recast the parquet equations for $F$ into a {formally} equivalent set of equations for $\Phi^\firr$.
Starting from the traditional parquet equations~\cite{Dominicis64,Dominicis64-2,Bickers04,Rohringer12}
for dual fermions~\cite{Astretsov19,Astleithner20}, we derive in Appendix~\ref{app:relationtoparquet}
the following {parquet expressions,
which could be interpreted like a set of parquet equations for the residual vertex},
\begin{widetext}
\begin{subequations}
\begin{align}
{\Phi}^{\firr,\ch}_{kk'q}=\;&\varphi^{\firr,\ch}_{\nu\nu'\omega}+{M}^{ph,\ch}_{kk'q}
-\frac{1}{2}{M}^{ph,\ch}_{k,k+q,k'-k}-\frac{3}{2}{M}^{ph,\sz}_{k,k+q,k'-k}+\frac{1}{2}{M}^{pp,\sing}_{kk',k+k'+q}
\;+\frac{3}{2}{M}^{pp,\trip}_{kk',k+k'+q}\label{eq:parquet_ch},\\
{\Phi}^{\firr,\sz}_{kk'q}=\;&\varphi^{\firr,\sz}_{\nu\nu'\omega}+{M}^{ph,\sz}_{kk'q}
-\frac{1}{2}{M}^{ph,\ch}_{k,k+q,k'-k}+\frac{1}{2}{M}^{ph,\sz}_{k,k+q,k'-k}-\frac{1}{2}{M}^{pp,\sing}_{kk',k+k'+q}
\;+\frac{1}{2}{M}^{pp,\trip}_{kk',k+k'+q}\label{eq:parquet_sp},\\
{\Phi}^{\firr,\sing}_{kk'q}\;\;=\;&\varphi^{\firr,\sing}_{\nu\nu'{\omega}}\;\,+{M}^{pp,\sing}_{kk'{q}}
\;+\frac{1}{2}{M}^{ph,\ch}_{kk',{q}-k'-k}-\frac{3}{2}{M}^{ph,\sz}_{kk',{q}-k'-k}  +\frac{1}{2}{M}^{ph,\ch}_{k,{q}-k',k'-k}-\frac{3}{2}{M}^{ph,\sz}_{k,{q}-k',k'-k},\label{eq:parquet_sing}\\
{\Phi}^{\firr,\trip}_{kk'q}\;\;=\;&\varphi^{\firr,\trip}_{\nu\nu'{\omega}}\;\,+{M}^{pp,\trip}_{kk'{q}}
\;+\frac{1}{2}{M}^{ph,\ch}_{kk',{q}-k'-k}+\frac{1}{2}{M}^{ph,\sz}_{kk',{q}-k'-k}  -\frac{1}{2}{M}^{ph,\ch}_{k,{q}-k',k'-k}-\frac{1}{2}{M}^{ph,\sz}_{k,{q}-k',k'-k}.\label{eq:parquet_trip}
\end{align}
\end{subequations}
\end{widetext}
Here, the labels $\ch,\sz,\sing,\trip$ denote the charge, spin, singlet, and triplet channels, respectively.
On the right-hand-side, $\varphi^\firr$ denotes the {local} analog to $\Phi^\firr$ {computed} from the AIM, see~Sec.~\ref{sec:aim}. This vertex plays a similar role as the fully irreducible vertex of the traditional
parquet equations (cf. Appendix~\ref{app:relationtoparquet}),
which, in the parquet approximation, is given by the \textit{bare} dual fermion interaction $f$ [cf. Eq.~\eqref{eq:dualaction}].
{It is important to remark that} Eqs.~\eqref{eq:parquet_ch}-\eqref{eq:parquet_trip}
represent the parquet {expression} for the residual vertex $\Phi^\firr$.
{Hence, they} are \textit{fully equivalent} to the parquet approximation for dual fermions.
{In spite of its analytical equivalence to the usual parquet expressions,
the formulation used here differs from the perspective of the numerical implementation.
In fact, in our BEPS method the starting point is represented by the corresponding
residual vertex $\varphi^\firr$ of the AIM.}

\begin{figure}[b]
\begin{center}
\includegraphics[width=0.4\textwidth]{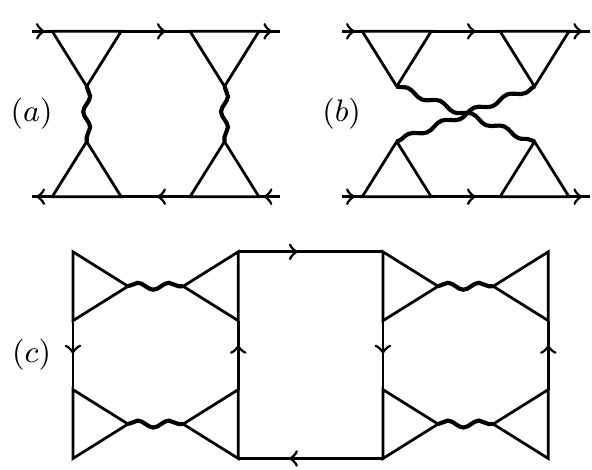}
\end{center}
    \caption{\label{fig:aslamazov}
    Multi-boson exchange generated by the ladder Eqs.~\eqref{eq:mbe_ladder_ph} and~\eqref{eq:mbe_ladder_pp}.
    Two-boson (Aslamazov-Larkin) exchange in particle-hole (a) and particle-particle (b) channels
    arises from contribution of SBE vertex $\Delta$ to the ladder kernel $S$ in
    Eqs.~\eqref{eq:ph_kernel_ch}-\eqref{eq:pp_kernel_trip}.
    (c) Higher multi-boson exchange due to mixing of vertical and horizontal particle-hole channels,
    origin is the feedback of MBE vertex $M$ on $S$.
    In this figure appropriate flavor labels and prefactors are omitted.
    }
    \end{figure}

{To further explicate the BEPS formalism, one should also note that} the vertex $M$ on the right-hand-side
plays essentially the role of the reducible vertex of the traditional parquet formalism.
The main difference is, however, that single-boson exchange diagrams are excluded from $M$.
Therefore, $M$ can be regarded as a vertex which {describes the
\textit{multi}-boson exchange (MBE, cf. Fig.~\ref{fig:aslamazov}) processes}.
To evaluate it {in practice}, we require an {\sl analog} to the {\sl Bethe-Salpeter} equations {(BSE)},
{which in the conventional formalism identifies the different scattering-channels through a separation of the two-particle reducible processes in the corresponding sectors.}

To this end, we define an auxiliary vertex $T$, which represents boson exchange {processes} of all orders
in a given channel. {Similarly to the BSE in the conventional formalism,}
it is given in terms of the ladder equations,

\begin{align}
{T}^{ph,\alpha}_{kk'q}=&{S}^{ph,\alpha}_{kk'q}
+\sum_{k''}{S}^{ph,\alpha}_{kk''q}G_{k''}G_{k''+q}{T}^{ph,\alpha}_{k''k'q}\notag\\
=&{S}^{ph,\alpha}_{kk'q}+{M}^{ph,\alpha}_{kk'q}
\label{eq:mbe_ladder_ph},
\end{align}
for the particle-hole channels ($\alpha=\ch,\sz$) and
\begin{align}
{T}^{pp,\delta}_{kk'q}=&{S}^{pp,\delta}_{kk'q}
\mp\frac{1}{2}\sum_{k''}{S}^{pp,\delta}_{kk''q}G_{k''}G_{q-k''}{T}^{pp,\delta}_{k''k'q}\notag\\
=&{S}^{pp,\delta}_{kk'q}+{M}^{pp,\delta}_{kk'q},
\label{eq:mbe_ladder_pp}
\end{align}
for the particle-particle channels ($\delta=\sing,\trip$), where $S$ denotes the respective ladder kernel.
{Note that} the vertex $T$ itself is not of interest here and need not be evaluated. Instead,
Eqs.~\eqref{eq:mbe_ladder_ph} and~\eqref{eq:mbe_ladder_pp} serve to evaluate all ladder diagrams
starting from the second order, that is, the vertex $M$. The ladder kernel is defined as follows,
\begin{widetext}
\begin{subequations}
\begin{align}
{S}^{ph,\ch}_{kk'q}=\,&{\Phi}^{\firr,\ch}_{kk'q}-{M}^{ph,\ch}_{kk'q}
-\frac{1}{2}{\Delta}^{ph,\ch}_{k,k+q,k'-k}-\frac{3}{2}{\Delta}^{ph,\sz}_{k,k+q,k'-k}
+\frac{1}{2}{\Delta}^{{pp},\sing}_{kk',q+k+k'}-2U^\ch,\label{eq:ph_kernel_ch}\\
{S}^{ph,\sz}_{kk'q}=\,&{\Phi}^{\firr,\sz}_{kk'q}-{M}^{ph,\sz}_{kk'q}
-\frac{1}{2}{\Delta}^{ph,\ch}_{k,k+q,k'-k}+\frac{1}{2}{\Delta}^{ph,\sz}_{k,k+q,k'-k}
-\frac{1}{2}{\Delta}^{{pp},\sing}_{kk',q+k+k'}-2U^\sz,\label{eq:ph_kernel_sp}\\
{S}^{pp,\sing}_{kk'{q}}\;\;=\,&{\Phi}^{\firr,\sing}_{kk'q}\;\,-{M}^{pp,\sing}_{kk'q}
\;\,+\frac{1}{2}{\Delta}^{ph,\ch}_{kk',{q}-k'-k}-\frac{3}{2}{\Delta}^{ph,\sz}_{kk',{q}-k'-k}
+\frac{1}{2}{\Delta}^{{ph},\ch}_{k,q-k',k'-k}-\frac{3}{2}{\Delta}^{{ph},\sz}_{k,q-k',k'-k}-U^\ch+3U^\sz,\label{eq:pp_kernel_sing}\\
{S}^{pp,\trip}_{kk'{q}}\;\;=\,&{\Phi}^{\firr,\trip}_{kk'q}\;\,-{M}^{pp,\trip}_{kk'q}
\;\,+\frac{1}{2}{\Delta}^{ph,\ch}_{kk',{q}-k'-k}+\frac{1}{2}{\Delta}^{ph,\sz}_{kk',{q}-k'-k}
-\frac{1}{2}{\Delta}^{{ph},\ch}_{k,q-k',k'-k}-\frac{1}{2}{\Delta}^{{ph},\sz}_{k,q-k',k'-k}.\label{eq:pp_kernel_trip}
\end{align}
\end{subequations}
\end{widetext}

Here finally the SBE vertex $\Delta$, which was introduced in Sec.~\ref{sec:strategy}, enters the parquet equations.
Further, by comparison with Eqs.~\eqref{eq:parquet_ch}-\eqref{eq:parquet_trip} 
one sees that also the multi-boson exchange represented by $M$ contributes to the kernel.
Ladder diagrams generated by $\Delta$ and $M$ are shown in Fig.~\ref{fig:aslamazov}.
{Although it may not be true in general,
we observed in our numerical applications that $\Delta$
yields the dominant contribution to the kernel $S$.}
{In these cases}, $S$ can be considered to mainly represent single-boson exchange,
while the contribution of $M$, {that is, the feedback of the multi-boson exchange on the kernel,
is} required to retain the exact equivalence to the
parquet approximation for dual fermions (see Appendix~\ref{app:relationtoparquet}).

For given vertices $\varphi^\firr$, $\Delta$ and Green's function $G$ the vertices 
$M$ and $\Phi^\firr$ in Eqs.~\eqref{eq:parquet_ch}-\eqref{eq:pp_kernel_trip} can be determined self-consistently.
One advantage of this calculation scheme is that $\Phi^\firr$ and $M$ decay at high frequencies.
Combined with the asymptotics of the dual propagator $G\propto\frac{1}{\nu^2}$
this leads to a rapid decay of Matsubara summations~\footnote{
A similar idea was used in Ref.~\cite{Krien19} to improve the feasibility of the DMFT susceptibility.}.
It is not necessary to take vertex asymptotics into account~\cite{Wentzell2020,Li16}.
Furthermore, the spatial dependence of the residual vertex $\Phi^\firr$ is short-ranged
compared to the full vertex $F$, which we exploit in Sec.~\ref{sec:tups} for a truncated unity approximation.

\subsection{Diagrammatic building blocks}
As in the traditional parquet formalism the Green's function is dressed with a
self-energy $\Sigma$, which can be calculated using the Schwinger-Dyson Eq.~\eqref{eq:sigma}
[where the full vertex is given via Eq.~\eqref{eq:jib_simple}].

However, the parquet equations for the residual vertex $\Phi^\firr$ in Sec.~\ref{sec:p3} also require
further prescriptions to calculate the fermion-boson coupling $\Lambda$ and the screened interaction $W$,
which are used to form the SBE vertices $\Delta$ in Eqs.~\eqref{eq:nabla_simple} and~\eqref{eq:nabla_simple_pp}.
The fermion-boson coupling is a three-leg vertex which does not contain insertions of the
bare Hubbard interaction $U$, see also Ref.~\cite{Krien19}.
We obtain it by removing the SBE vertex $\Delta^{ph}$
from the full vertex $F$ and attaching two {(dual)} Green's functions.
We begin with the charge and spin channels ($\alpha=\ch,\sz$),
\begin{align}
  \Lambda^{\alpha}_{kq} = \lambda^\alpha_{\nu\omega}+\sum_{k'}(F^\alpha-\Delta^{ph,\alpha})_{kk'q}G_{k'}G_{k'+q}
  \lambda^\alpha_{\nu'\omega}.\label{eq:hedin_chsp}
\end{align}
This equation highlights a peculiarity of bosonic correlation functions
in the dual fermion approach (see Appendix~\ref{app:susc}):
Whenever we form a bosonic end-point of a dual fermion diagram using two Green's functions,
we also attach the impurity vertex $\lambda$.
As a result, the leading contribution to $\Lambda$ is not simply $1$, as for lattice fermions~\cite{Krien19-3},
but it is the fermion-boson coupling $\lambda$ of the impurity,
which {is defined by all corresponding fully local diagrams of the auxiliary AIM~\cite{vanLoon14,vanLoon15}.}

\begin{figure}
\begin{center}
\includegraphics[width=0.45\textwidth]{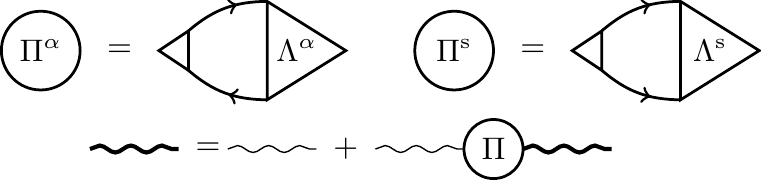}
\end{center}
    \caption{\label{fig:polarization}
    Top: Polarization for particle-hole (left, $\alpha=\ch,\sz$) and singlet particle-particle channel (right).
    Small triangles denote the fermion-boson coupling $\lambda$ of the AIM.
    Bottom: Dyson equation for the screened interaction $W$ (thick wiggly),
    thin lines denote the screened interaction $w$ of the AIM.
    }
    \end{figure}

Next, we write the screened interaction as
\begin{align}
W^\alpha(q)=\frac{w^\alpha(\omega)}{1-w^\alpha(\omega)\Pi^\alpha(q)},\label{eq:dyson_boson_ph}
\end{align}
where $w^\alpha(\omega)$ is the screened interaction of the AIM defined in Eq.~\eqref{eq:w}
and $\Pi$ is the dual polarization function,
\begin{align}
\Pi^\alpha(q)=\sum_k\lambda^\alpha_{\nu\omega}G_{k}G_{k+q}\Lambda^{\alpha}_{kq},\label{eq:polarization}
\end{align}
which is shown as a diagram in Fig.~\ref{fig:polarization}.
Again, to form the second bosonic end-point of the polarization,
we attached the vertex $\lambda$, this time from the left.

So far, we have discussed the particle-hole channels $\alpha=\ch,\sz$.
However, the bare Hubbard interaction also couples to a singlet particle-particle channel, $\alpha=\sing$.
In this channel the fermion-boson coupling takes the form~\footnote{
The minus sign for the impurity vertex $\lambda^\sing$ in Eq.~\eqref{eq:hedin_singlet} is plausible because
Eq.~\eqref{eq:polarization_singlet} is quadratic in $\lambda^\sing$.
The latter is given to leading order by $-1$~\cite{Krien19-2}, so Eq.~\eqref{eq:hedin_singlet}
leads to an overall minus sign for the leading order of $\Pi$, as expected for the singlet channel~\cite{Krien19-3}.},
\begin{align}
{\Lambda}^{\sing}_{k{q}}\!=\!-\lambda^\sing_{\nu{\omega}}
+\frac{1}{2}\sum_{k'}(F^\sing\!-\!{\Delta}^{pp,\sing})_{kk'{q}}
{G}_{k'}{G}_{{q}-k'}\lambda^\sing_{\nu'{\omega}},\label{eq:hedin_singlet}
\end{align}
where $F^\sing$ is the singlet vertex function~\footnote{
$F^\sing_{kk'q}=\frac{1}{2}F^\ch_{kk',q-k-k'}-\frac{3}{2}F^\sz_{kk',q-k-k'}$}.
The reducible vertex ${\Delta}^{pp,\sing}$ for this channel is defined in Eq.~\eqref{eq:nabla_simple_pp},
where the corresponding screened interaction reads,
\begin{align}
W^\sing(q)=\frac{w^\sing(\omega)}{1-\frac{1}{2}w^\sing(\omega)\Pi^\sing(q)},\label{eq:dyson_boson_pp}
\end{align}
and the polarization is given as (see also Fig.~\ref{fig:polarization}),
\begin{align}
\Pi^\sing(q)=\sum_k\lambda^\sing_{\nu\omega}G_{k}G_{q-k}\Lambda^{\sing}_{kq}.\label{eq:polarization_singlet}
\end{align}
All quantities in this section are defined for dual fermions.
The prescription for the renormalization of the fermion-boson coupling in Eqs.~\eqref{eq:hedin_chsp}
and~\eqref{eq:hedin_singlet} is the dual fermion analog to the method introduced
in Ref.~\cite{Krien19-3} for lattice fermions.

\subsection{Truncated unity approximation}\label{sec:tups}
The parquet expressions for the residual vertex $\Phi^\firr$ in Sec.~\ref{sec:p3}
improve the feasibility {of the parquet approximation for dual fermions.}
Nonetheless, similarly as in the standard parquet implementations,
the vertices quickly become very large with increasing lattice size~\cite{Kauch19}.
To {mitigate} this problem, Ref.~\cite{Eckhardt20} introduced a truncated unity parquet solver (TUPS),
using a form-factor expansion of the various vertex functions.
In {the same} spirit, we {can further} improve the feasibility by transforming the ladder
equations~\eqref{eq:mbe_ladder_ph} and~\eqref{eq:mbe_ladder_pp}
into the form-factor basis as in Eq.~\eqref{eq:ff}, for example,
\begin{align}
{T}^{ph,\alpha}_{\ell\ell'q}=&{S}^{ph,\alpha}_{\ell\ell'q}
+\sum_{\ell_1\ell_2}{S}^{ph,\alpha}_{\ell\ell_1q}X^0_{\ell_1\ell_2q}{T}^{ph,\alpha}_{\ell_2\ell'q},
\label{eq:mbe_ladder_ph_ff}
\end{align}
where $X^0_{\ell\ell'q}$ is a dual particle-hole bubble in the form-factor basis.
The expansion is then truncated at a number $N_\ell$ of form factors (see Sec.~\ref{sec:strategy}).
In the (truncated) form-factor basis it is feasible to solve the
ladder equation~\eqref{eq:mbe_ladder_ph_ff} by inversion,
which may improve the convergence of the parquet solver
compared to previous implementations which build the ladder diagrams iteratively~\cite{Tam13,Kauch19}.

On the other hand, we keep the full momentum-dependence of the fermion-boson coupling $\Lambda(k,q)$.
Therefore, to evaluate Eqs.~\eqref{eq:hedin_chsp} and~\eqref{eq:hedin_singlet},
we obtain the vertex $M$ from the back-transformation,
\begin{align}
M(k,k',q)=\sum_{\lv\lv'}\psi(\lv,\kv)M(\ell,\ell',q)\psi(\lv',\kv').\label{eq:ff_phi}
\end{align}
In the implementation only $M(\ell,\ell',q)$ is stored and Eq.~\eqref{eq:ff_phi} is used when $M(k,k',q)$ is needed.
The calculation of the ladder kernel $S$ in Eqs.~\eqref{eq:ph_kernel_ch}-\eqref{eq:pp_kernel_trip} requires momentum shifts,
Ref.~\cite{Eckhardt20} describes in detail how they can be handled in the form-factor basis,
see also Appendix~\ref{app:formf}.
{Of course,} the momentum shifts imply a (truncated-unity) cutoff
with respect to all three momenta~\cite{Eckhardt20}.
{In our scheme, however,} this problem is alleviated because the
truncation does not affect the single-boson exchange $\Delta$.

\begin{figure}
    \begin{center}
        \includegraphics[width=0.55\textwidth]{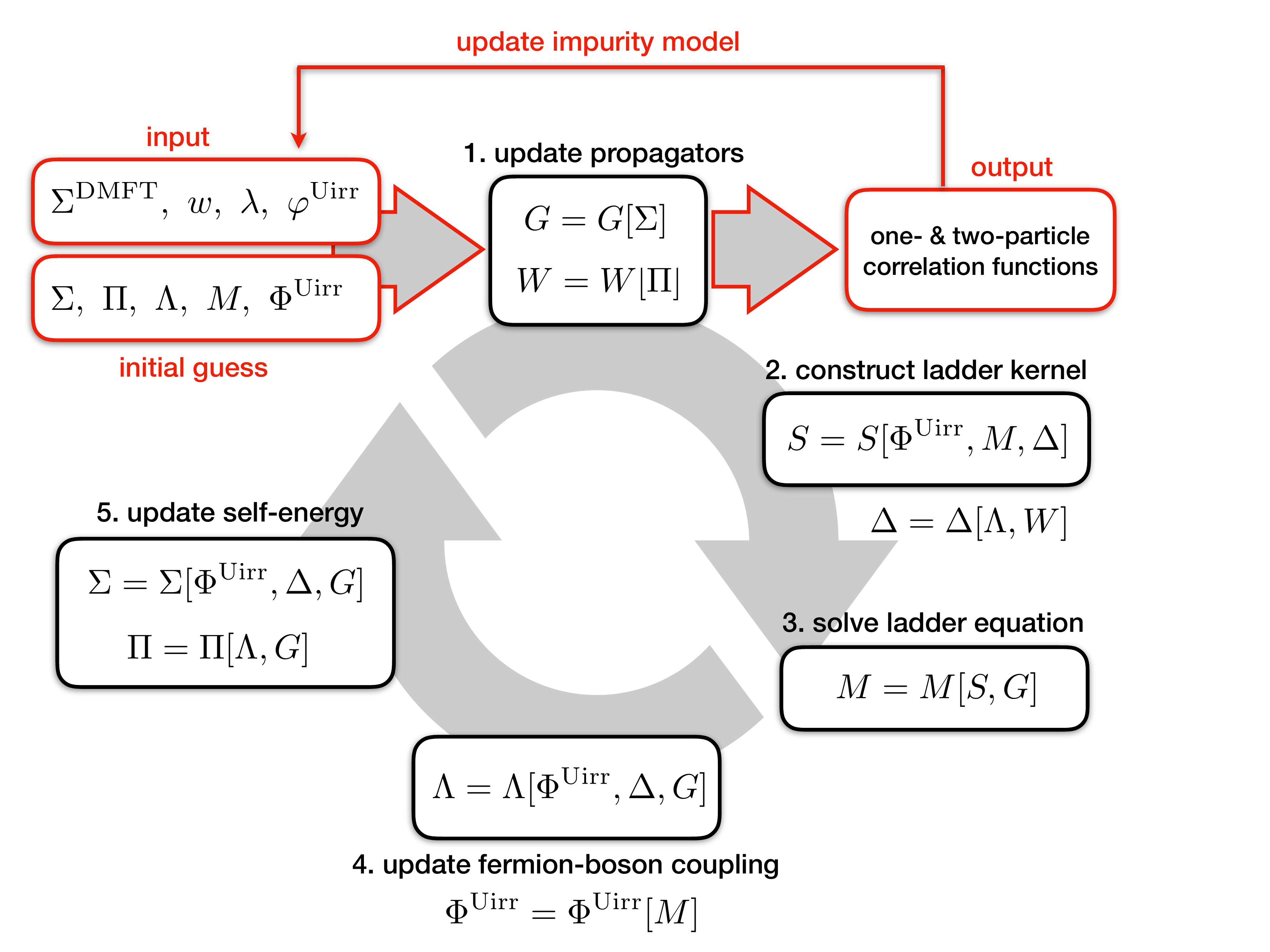}
    \end{center}
    \vspace{-0.5cm}
    \caption{\label{fig:cycle} (Color online) 
    Self-consistent cycle of the BEPS method. 
    Highlighted (in red) are the input and output 
    as well as the external self-consistency cycle to update the AIM.
    }
\end{figure}

\subsection{Calculation cycle}\label{sec:cycle}
Fig.~\ref{fig:cycle} shows the calculation cycle of the BEPS method.

\noindent \textit{Step 0: Initial guess.} 
The calculation begins with the solution of the AIM~\eqref{eq:aim} to obtain the impurity correlation functions.
For an agnostic guess we set $\Sigma=0, \Lambda=\lambda, \Phi^\firr=\varphi^\firr, M=0$,
the corresponding polarization $\Pi$ is given via Eqs.~\eqref{eq:polarization},~\eqref{eq:polarization_singlet}.
To start closer to the solution, or near an instability, we can use the output of a previous BEPS calculation.

\noindent \textit{Step 1: Update propagators.} 
The fermionic and bosonic propagators $G$ and $W$ are updated using the Dyson equations~\eqref{eq:dyson},~\eqref{eq:dyson_boson_ph}, and~\eqref{eq:dyson_boson_pp}. 

\noindent \textit{Step 2: Construct ladder kernel.}
The kernel $S$ is built from Eqs.~\eqref{eq:ph_kernel_ch}-\eqref{eq:pp_kernel_trip}
{[where the vertices $\Delta$ are given by Eqs.~\eqref{eq:nabla_simple} and~\eqref{eq:nabla_simple_pp}]}
and transformed to the form-factor basis (see Appendix~\ref{app:formf}).

\noindent \textit{Step 3: Solve ladder equations.}
MBE vertices $M$ are obtained via inversion of Eqs.~\eqref{eq:mbe_ladder_ph} and~\eqref{eq:mbe_ladder_pp}.

\noindent \textit{Step 4: Update fermion-boson coupling.}
$\Lambda$ is updated via Eqs.~\eqref{eq:hedin_chsp} and~\eqref{eq:hedin_singlet}.
In these equations, the full vertex $F$ is given by the SBE decomposition in Eq.~\eqref{eq:jib_simple}.
The residual vertex $\Phi^\firr$ is obtained from the MBE vertices $M$
via the parquet equations~\eqref{eq:parquet_ch}-\eqref{eq:parquet_sing}
and back-transformation to the momentum-basis as in Eq.~\eqref{eq:ff_phi}
[momentum-shifts are treated as in Appendix~\ref{app:formf}].

\noindent \textit{Step 5: Update self-energies.} The self-energy $\Sigma$ and the polarization $\Pi$
are calculated from Eqs.~\eqref{eq:sigma}, \eqref{eq:polarization},
and \eqref{eq:polarization_singlet}, respectively. 
In Eq.~\eqref{eq:sigma} the full vertex is given as described in Step 4~\footnote{
In the calculation of the self-energy via Eq.~\eqref{eq:sigma} it is convenient to bring
all vertex components into their channel-native form~\cite{Eckhardt20}.
}.

\noindent Steps from 1 to 5 are iterated until convergence.
Optionally, the hybridization function $h_\nu$ of the AIM~\eqref{eq:aim}
is updated (outer self-consistency) and the cycle is restarted from Step 0
(this work: $h\equiv h^\text{DMFT}$).

\begin{figure}
    \begin{center}
        \includegraphics[width=0.53\textwidth]{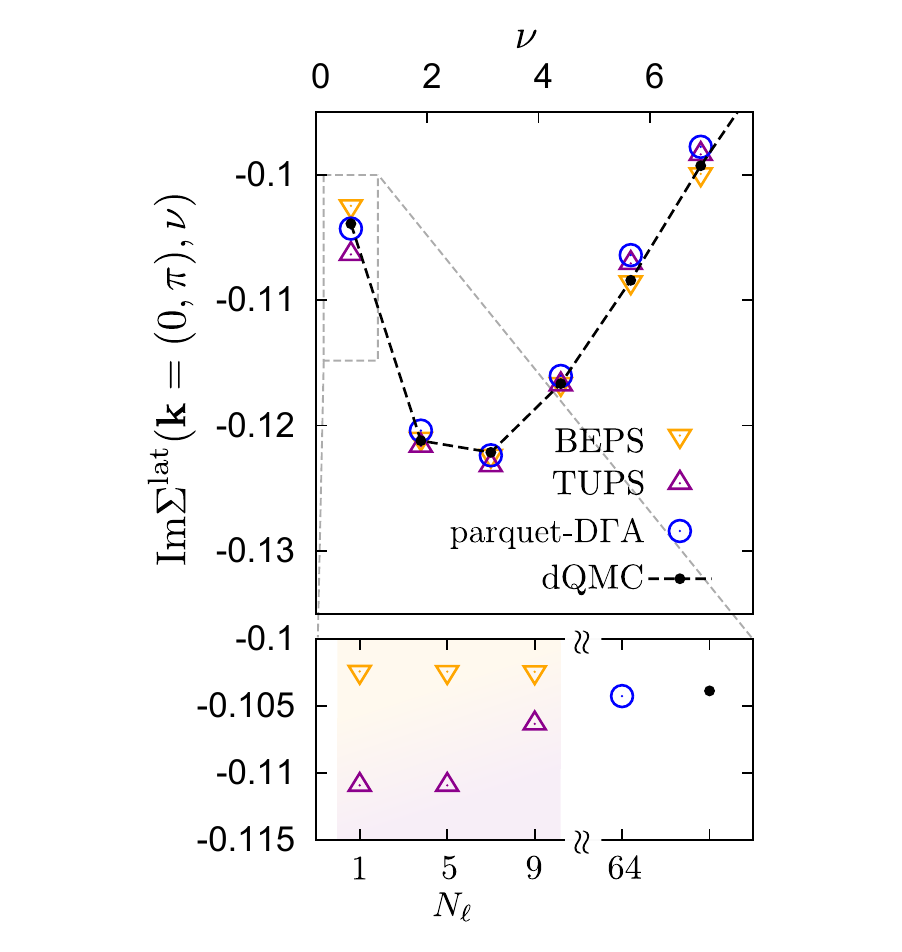}
    \end{center}
    \vspace{-0.5cm}
    \caption{\label{fig:dga}
Self-energy at the anti-nodal point at $U/t=2$ and $T/t=0.2$.
Top: BEPS self-energy and TUPS-D$\Gamma$A using $8\times8$ lattice and $N_\ell=9$ form-factors, respectively,
compared to the dQMC data of Ref.~\cite{Hille20}.
Circles represent the untruncated ($N_\ell=64$) parquet D$\Gamma$A result.
Bottom: Convergence of BEPS and TUPS-D$\Gamma$A with the form-factors.
    }
\end{figure}

\subsection{Implementation notes}
Our implementation of the BEPS method is a working prototype based on the C++ libraries of the
ladder dual fermion/boson implementation of H. Hafermann
and E.G.C.P. van Loon~\cite{Hafermann09,vanLoon14}, but the alterations to the code are substantial.
For the truncated unity approximation we use an implementation of the form-factors for the square
lattice by C. Eckhardt~\cite{Eckhardt18,Eckhardtformf}.

The numerical effort of Eqs.~\eqref{eq:hedin_chsp} and~\eqref{eq:hedin_singlet} is discussed
in Ref.~\cite{Krien19-3}, {corresponding to $\propto N_\nu^2N_k^2N_\omega N_q$ floating point operations.}
The most expensive step at each iteration is the transformation of the ladder kernel $S$
in Eqs.~\eqref{eq:ph_kernel_ch}-\eqref{eq:pp_kernel_trip}
to the form-factor basis [cf. Eqs.~\eqref{eq:ff} and~\eqref{eq:ff_phi}],
which requires $\propto N_\ell^2 N_\nu^2 N_\omega N^2_k N_q$ floating point operations.
We use a parallel code where each process performs the transformation and solves the ladder
equations~\eqref{eq:mbe_ladder_ph} and~\eqref{eq:mbe_ladder_pp} for one momentum energy $q=(\qv,\omega)$,
but the numerical effort still scales $\propto N_\ell^2 N_\nu^2 N^2_k$ for each process.
In the applications we set the lattice size to $8\times8$ and $16\times16$ sites,
but a $32\times32$ lattice is feasible~\cite{Krien20-3}.
To further increase the lattice size it is appealing to port the implementation to GPUs~\cite{Astretsov19}.
The method is memory-efficient, indeed, the largest object stored 
during calculations is the fermion-boson vertex $\Lambda(k,q)$ {of size $N_\nu N_kN_\omega N_q$},
which is in turn split into $N_kN_\nu$ pieces, hence, each process handles only a vector of length $N_qN_\omega$.
Further, the numbers $N_\nu$ and $N_\omega$ of Matsubara frequencies
and the number $N_\ell$ of form-factors can be kept small compared to other schemes,
as discussed in Sec.~\ref{sec:p3}.

\begin{figure}
    \begin{center}
        \includegraphics[width=0.5\textwidth]{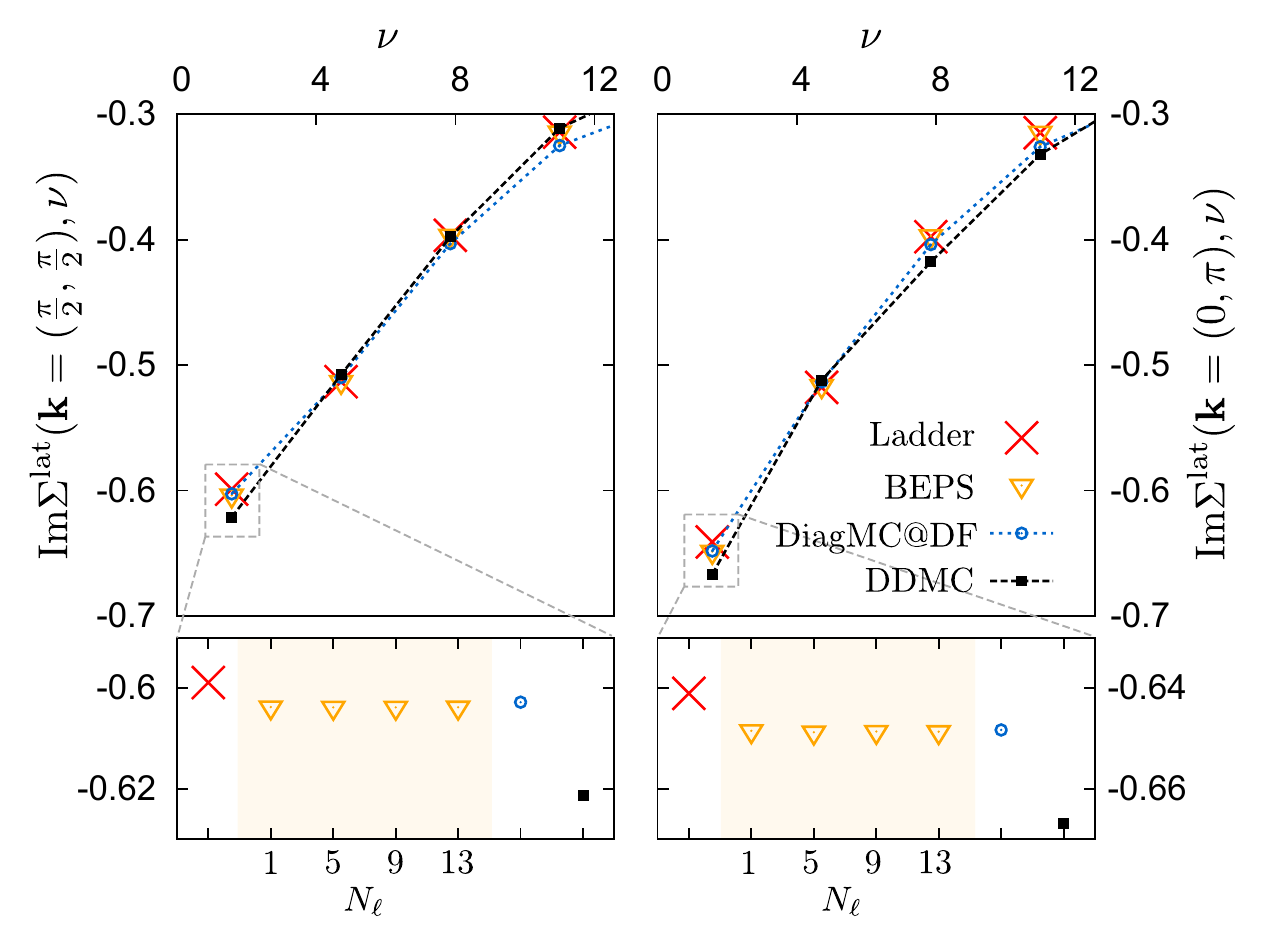}
    \end{center}
    \vspace{-0.5cm}
    \caption{\label{fig:u4b2}
     Self-energy at the nodal (left) and anti-nodal (right) points for $U/t=4$ and $T/t=0.5$.
     Triangles show the BEPS result for various cutoffs $N_\ell$ of the truncated unity
     (top panels: $N_\ell=13$).
     Full (dashed) black lines show DiagMC@DF (DDMC), crosses indicate the
     ladder dual fermion approximation. Bottom panels show a closeup of $\nu=\pi T$.
    }
\end{figure}

Several symmetries are used to improve the performance:
The point-group symmetry~\cite{Thomale13} implies that $\Lambda(k,q)$ is invariant when
we project the momentum $\kv$ into the irreducible Brillouin zone
and apply the same symmetry operation to $\qv$~\footnote{
In general only one of the momenta $\kv, \qv$ can be mapped to the irreducible Brillouin zone,
therefore, $\Lambda(k,q)$ needs to be stored for $N^\text{irr}_kN_q$ momenta.
}.
Time-reversal and SU($2$) symmetry~\cite{Rohringer12,Krien19-3} imply $S(\ell,\ell',q)=S(\ell',\ell,q)$
for the expensive ladder kernel and we evaluate only a triangle of this matrix.

\section{Benchmarks at half-filling}\label{sec:benchmark}
We apply the BEPS method to the half-filled Hubbard model~\eqref{eq:hubbardmodel}
with nearest-neighbor hopping, interaction $U/t=2,4,8$ and temperatures $T/t=0.5$ and $T/t=0.2$.
The lattice size corresponds to $8\times8$ sites at $T/t=0.5$ and $16\times16$ sites at $T/t=0.2$.
The Matsubara cutoff for Eqs.~\eqref{eq:hedin_chsp} and~\eqref{eq:hedin_singlet} is $N_\nu=N_\omega=14$.
The ladder equations~\eqref{eq:mbe_ladder_ph} and~\eqref{eq:mbe_ladder_pp}
are evaluated using $N_\nu=8$ fermionic frequencies.
Appendix~\ref{app:nu_convergence} shows an example for the frequency convergence of BEPS.
We use $1\leq N_\ell\leq13$ form-factors.

\subsection{Lattice self-energy at weak coupling}
We begin with a quantitative comparison of the lattice self-energy~\eqref{eq:sigma_lat}
with results from the literature for weak coupling $U/t=2$ and temperature $T/t=0.2$.
Here, Ref.~\cite{Eckhardt20} recently reported results from the parquet dynamical vertex approximation
(parquet D$\Gamma$A) and compared them to the truncated unity approximation (TUPS-D$\Gamma$A).
As a numerically exact reference we use a determinant quantum Monte Carlo
(dQMC,~\cite{Scalapino81}) result of Ref.~\cite{Hille20}.
The top panel of Fig.~\ref{fig:dga} shows a good agreement of BEPS with both dQMC and D$\Gamma$A.

\begin{figure}
    \begin{center}
         \includegraphics[width=0.45\textwidth]{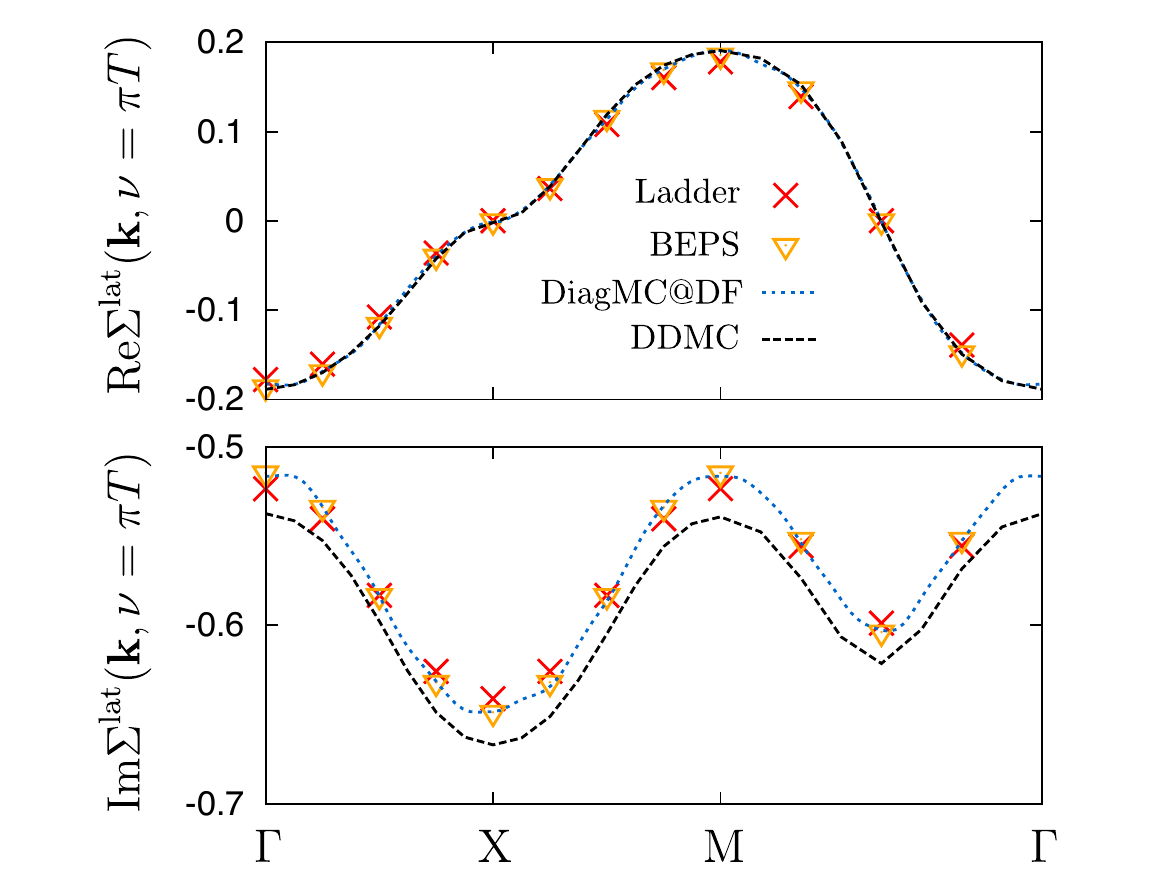}
    \end{center}
    \vspace{-0.5cm}
    \caption{\label{fig:u4b2_path}
     $U/t=4, T/t=0.5$. Real and imaginary part of the self-energy in the Brillouin zone
     at the first Matsubara frequency $\nu=\pi T$. BEPS self-energy shown for $N_\ell=13$.
    }
\end{figure}

The bottom panel of Fig.~\ref{fig:dga} shows the convergence of BEPS and TUPS-D$\Gamma$A
with the number of form factors $N_\ell$.
As explained in the previous sections, the working hypothesis of the BEPS method is
that it is beneficial to use the truncated unity approximation only for the residual vertex
$\Phi^\firr$ of the SBE decomposition~\eqref{eq:jib_simple} because it {should}
lead  {--per construction--} to a fast convergence with the number of form factors. 
Indeed, in this regime BEPS essentially converges with only one form factor, $N_\ell=1$.
The slower convergence of TUPS-D$\Gamma$A compared to BEPS is a consequence of the different use of the
truncated unity approximation in these methods (see Sec.~\ref{sec:strategy}).

\subsection{Lattice self-energy at strong coupling}
Ref.~\cite{Gukelberger17} presented a stochastic sampling of dual fermion diagrams (DiagMC@DF),
{with the usual truncation of the effective three-particle interaction.}
%of the up to a given order {in the two-particle DF interaction.}
Supplemental material of the reference contains a comprehensive dataset,
also in comparison with numerically exact diagrammatic determinant Monte Carlo (DDMC,~\cite{Burovski06}).
This gives us the opportunity to compare the BEPS method over a wide parameter range,
in fact, one of the techniques used in Ref.~\cite{Gukelberger17} is numerically exact for lattice fermions (DDMC),
the other for dual fermions with a quartic interaction potential (DiagMC@DF).
Therefore, the DiagMC@DF data correspond to the target result,
provided it is converged with respect to the perturbation order.
We show the results of Ref.~\cite{Gukelberger17} corresponding to order $\mathcal{O}(6)$.

\begin{figure}
    \begin{center}
        \includegraphics[width=0.5\textwidth]{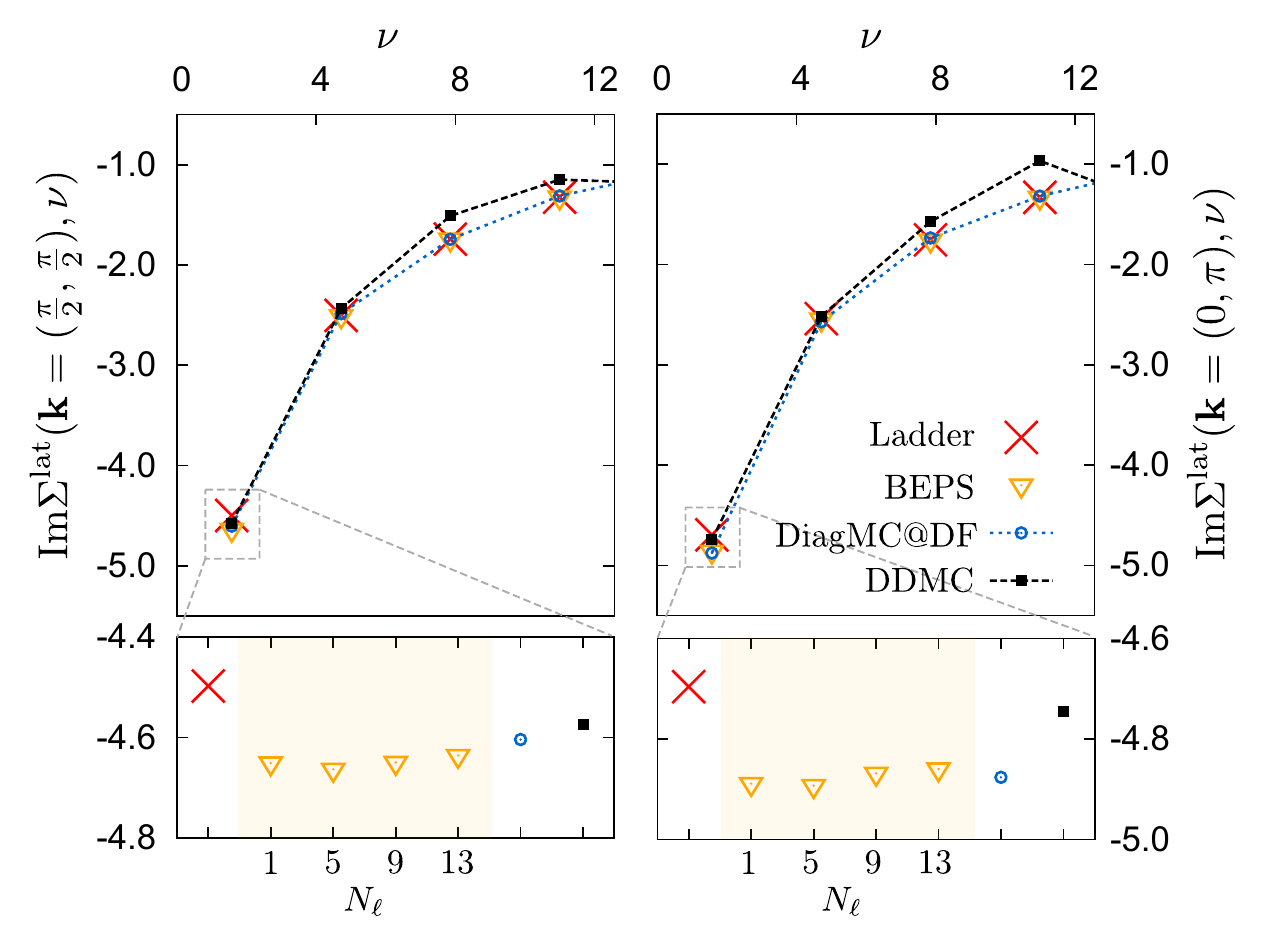}
    \end{center}
    \vspace{-0.5cm}
    \caption{\label{fig:u8b2}
     Self-energy at the nodal and anti-nodal points for $U/t=8$ and $T/t=0.5$.
     Labels as in Fig.~\ref{fig:u4b2}.}
\end{figure}

Fig.~\ref{fig:u4b2} shows the imaginary part of the self-energy at the antinodal and 
nodal points for $U/t=4$ and $T/t=0.5$.
The bottom panels show that the BEPS self-energy is again almost independent of the number
of form factors $1\leq N_\ell\leq13$.
As expected, the BEPS results lie closer to DiagMC@DF than the self-energy of the ladder dual fermion approach
(LDFA,~\cite{Hafermann09}).
Fig.~\ref{fig:u4b2_path} shows real and imaginary part of the self-energy
at the first Matsubara {frequency along the $\Gamma-X-M-\Gamma$ path} in the Brillouin zone.
In case of the real part, there is a good agreement between DDMC, DiagMC@DF and BEPS,
whereas for the imaginary part the dual fermion methods are
consistent with each other but show a small {low-frequency} offset compared to DDMC.
{This can be reasonably ascribed to} the truncation of
the dual fermion interaction after the quartic term~\cite{Gukelberger17}.

\begin{figure}
    \begin{center}
         \includegraphics[width=0.45\textwidth]{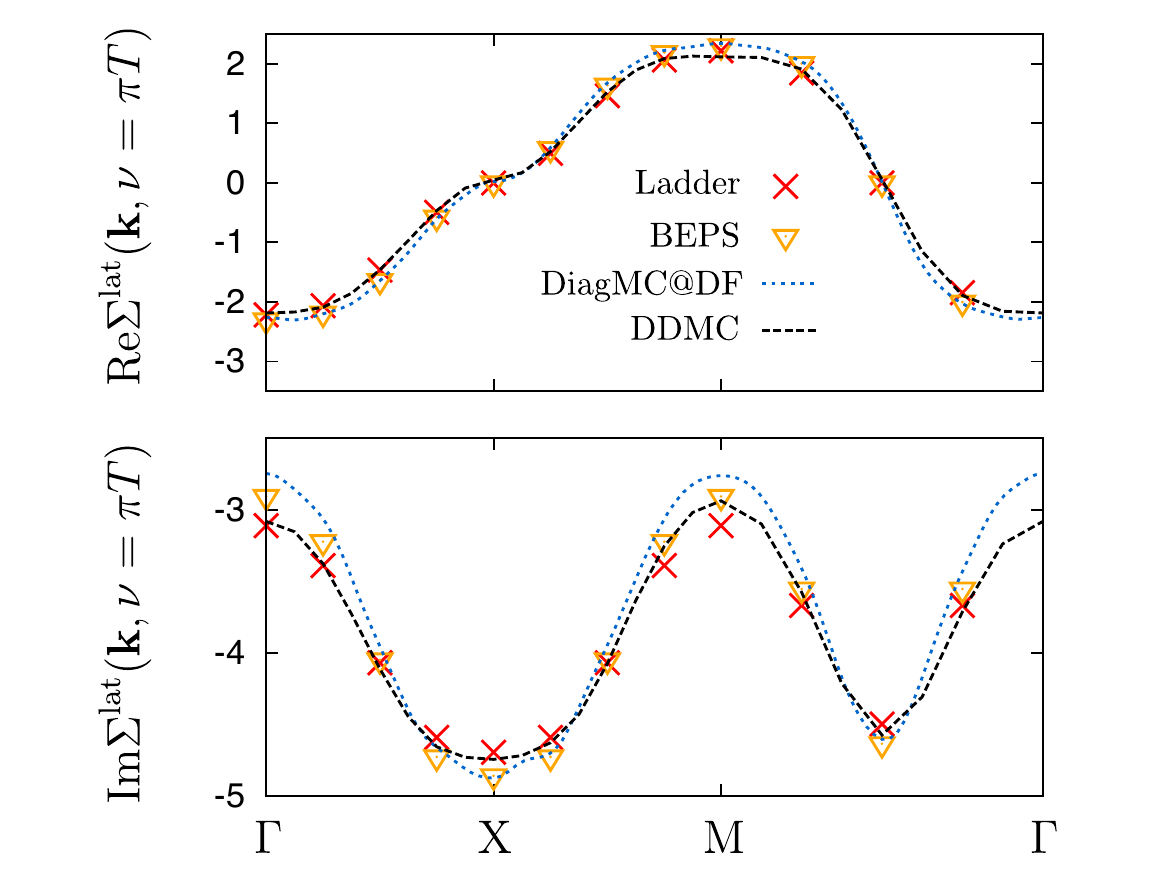}
    \end{center}
    \vspace{-0.5cm}
    \caption{\label{fig:u8b2_path}
     $U/t=8, T/t=0.5$, labels as in Fig.~\ref{fig:u4b2_path}.
     Notice that the DiagMC@DF result of Ref.~\cite{Gukelberger17} (dashed blue)
     is not fully converged in the expansion order.
    }
\end{figure}

We turn to the delicate regime $U/t=8$, see Figs.~\ref{fig:u8b2} and~\ref{fig:u8b2_path},
where for $T/t=0.5$ we find a slightly stronger dependence of
the BEPS result on the number of form factors.
At the node and antinode the results for different $N_\ell$ extrapolate accurately to DiagMC@DF,
see bottom panels of Fig.~\ref{fig:u8b2}.
Fig.~\ref{fig:u8b2_path} shows that in some parts of the Brillouin zone the BEPS
result lies closer to DDMC than to DiagMC@DF, however,
the latter is not fully converged with respect to the perturbation order~\cite{Gukelberger17}.

\begin{figure}
    \begin{center}
      \includegraphics[width=0.47\textwidth]{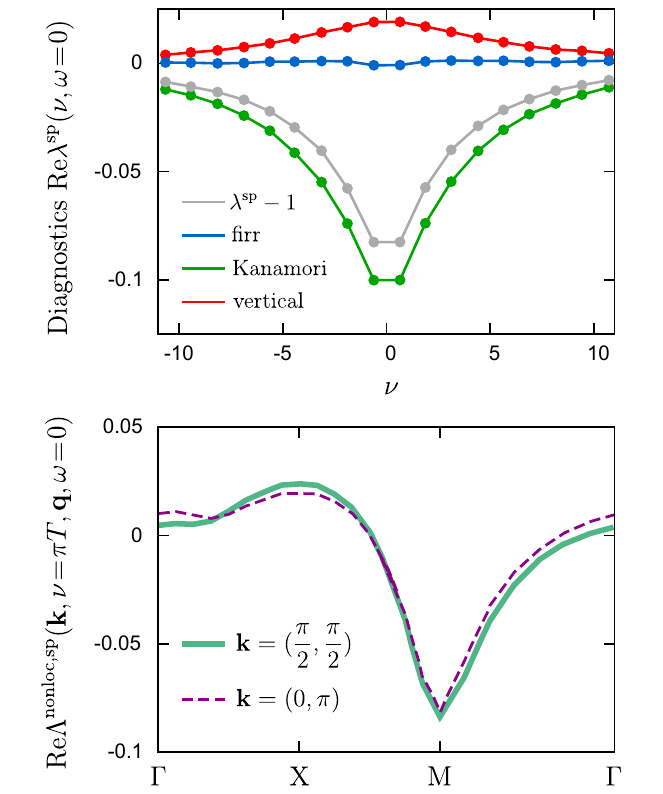}
    \end{center}
    \vspace{-0.5cm}
    \caption{\label{fig:lambdaimp} 
    Top: Local fermion-spinboson coupling $\lambda^\sz(\nu,\omega=0)$ (gray)
    for a DMFT calculation at half-filling, $U/t=2$ and $T/t=0.2$,
    corresponding to the BEPS self-energy in Fig.~\ref{fig:dga}.
    Colored lines show the vertex corrections which yield the frequency dependence
    of $\lambda^\sz$ (see text).
    Bottom: Nonlocal component at node and antinode as a function of $\qv$.
     }
\end{figure}

\subsection{Fermion-boson coupling}
We analyze a key quantity of BEPS, the fermion-boson coupling $\Lambda(k,q)$
defined in Eq.~\eqref{eq:hedin_chsp},
\begin{align}
  \Lambda(k,q)=\lambda(\nu,\omega)+\Lambda^\text{nonloc}(k,q).~\label{eq:lambdanonloc}
\end{align}
The hybridization of the AIM~\eqref{eq:aim} corresponds to the DMFT solution,
which provides the local vertex $\lambda$ in Eq.~\eqref{eq:hedinvertex},
and the BEPS method adds nonlocal corrections.
At half-filling $\lambda$ is real, $\Lambda^\text{nonloc}$ is in general complex.
We set $U/t=2, T/t=0.2$ and examine the coupling $\Lambda^\sz$ of fermions to spin fluctuations,
this vertex plays a role in the spin-fermion model~\cite{Schmalian99,Katanin09}.

We begin with the local component $\lambda^\sz(\nu,\omega=0)$
drawn in the top panel of Fig.~\ref{fig:lambdaimp},
which is suppressed for small $|\nu|$ compared to its non-interacting value $1$.
This effect is the result of particle-particle (Kanamori) screening~\cite{Kanamori63,Katanin09},
which can be seen explicitly by calculating the contribution
of singlet fluctuations to $\lambda^\sz$~\footnote{
The particle-particle screening is given as~\cite{Krien19-3}
   $\lambda^\text{Kanamori}_{\nu\omega}=-\frac{1}{2}\sum_{\nu'}(\nabla^{pp,\sing}_{\nu\nu',\omega+\nu+\nu'}-U^\sing)g_{\nu'}g_{\nu'+\omega}$.},
see green curve in the top panel of Fig.~\ref{fig:lambdaimp}.
The singlet fluctuations are given by the impurity SBE vertex $\nabla^{pp}$ in Eq.~\eqref{eq:jib_aim}.
The next largest vertex correction corresponds to an enhancement of $\lambda^\sz$
due to (vertical) spin and charge boson exchange (red),
$\nabla^{\overline{ph}}$, whereas the contribution of the (local) residual
vertex $\varphi^\firr$ is small in the considered regime (blue)~\footnote{
The horizontal SBE vertex $\nabla^{ph}$ does not contribute to $\lambda^\sz$,
which is (horizontally) irreducible with respect to the bare interaction~\cite{Krien19-3}.}.
As a result, DMFT provides a local Kanamori screening of fermions from spin fluctuations
as a starting point for the BEPS calculation.
One may note that our analysis of $\lambda^\sz$ corresponds, quite literally,
to a {\sl fluctuation diagnostic}~\cite{Gunnarsson15} of the fermion-boson coupling.

Next, we examine the nonlocal corrections, the bottom panel of Fig.~\ref{fig:lambdaimp} shows
$\Lambda^\text{nonloc,sp}(\kv,\nu=\pi T,\qv,\omega=0)$ where $\kv$ corresponds to the antinode or node
and the bosonic momentum $\qv$ runs along the high-symmetry path of the Brillouin zone.
Around $\qv=(\pi,\pi)$ the nonlocal component is negative,
corresponding to the screening of fermions from bosons with this momentum,
which is added to the Kanamori screening from the impurity model discussed above.
In the considered regime $\Lambda^\text{nonloc,sp}$ does not exhibit appreciable
differentiation with respect to the fermionic momentum $\kv$,
this occurs only at low temperature, in the pseudogap regime~\cite{Krien20-3}.

\section{Benchmarks away from half-filling}\label{sec:benchmark_doped}

\begin{figure}
    \begin{center}
         \includegraphics[width=0.48\textwidth]{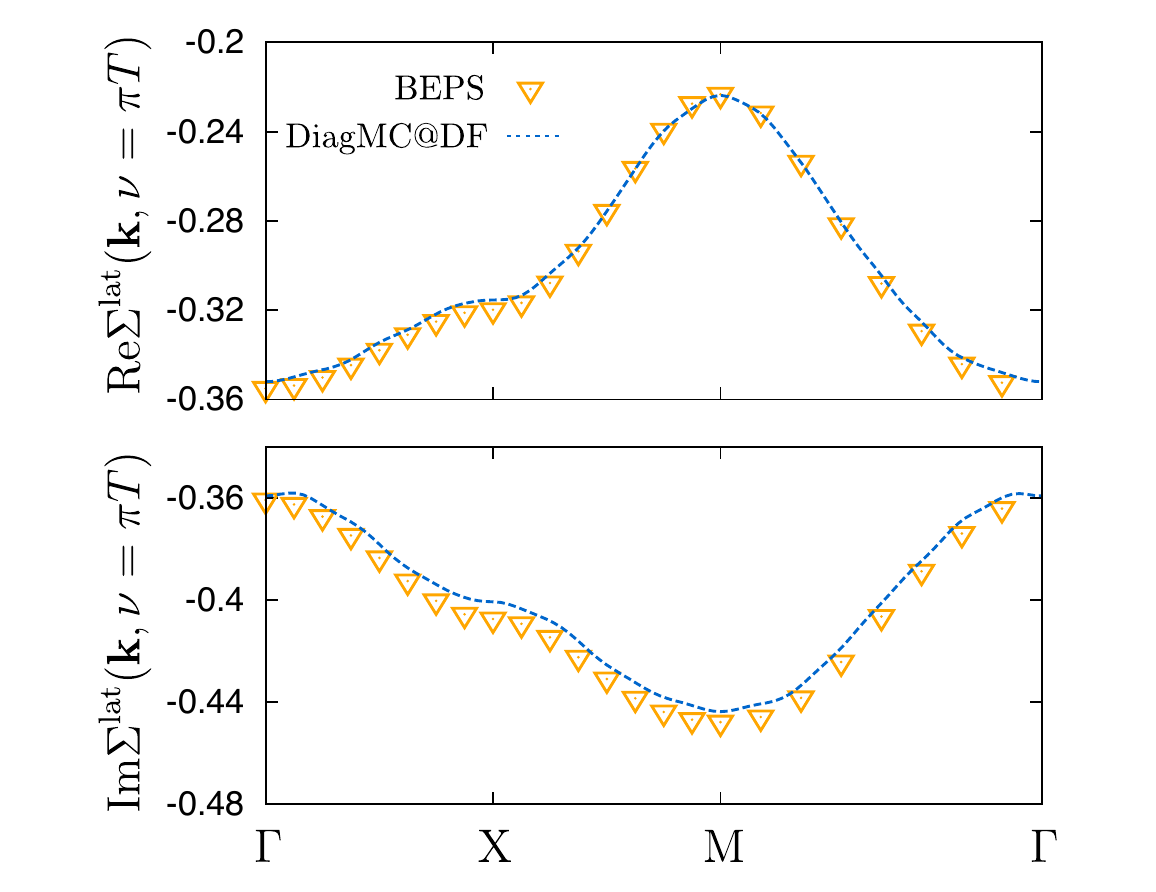}
    \end{center}
    \vspace{-0.5cm}
    \caption{\label{fig:u4b2_doped_path}
     $U/t=4, T/t=0.5, n\approx0.76$. 
     Real and imaginary part of the self-energy in
     the Brillouin zone at the first Matsubara frequency $\nu=\pi T$.
    }
\end{figure}

\begin{figure}
    \begin{center}
         \includegraphics[width=0.48\textwidth]{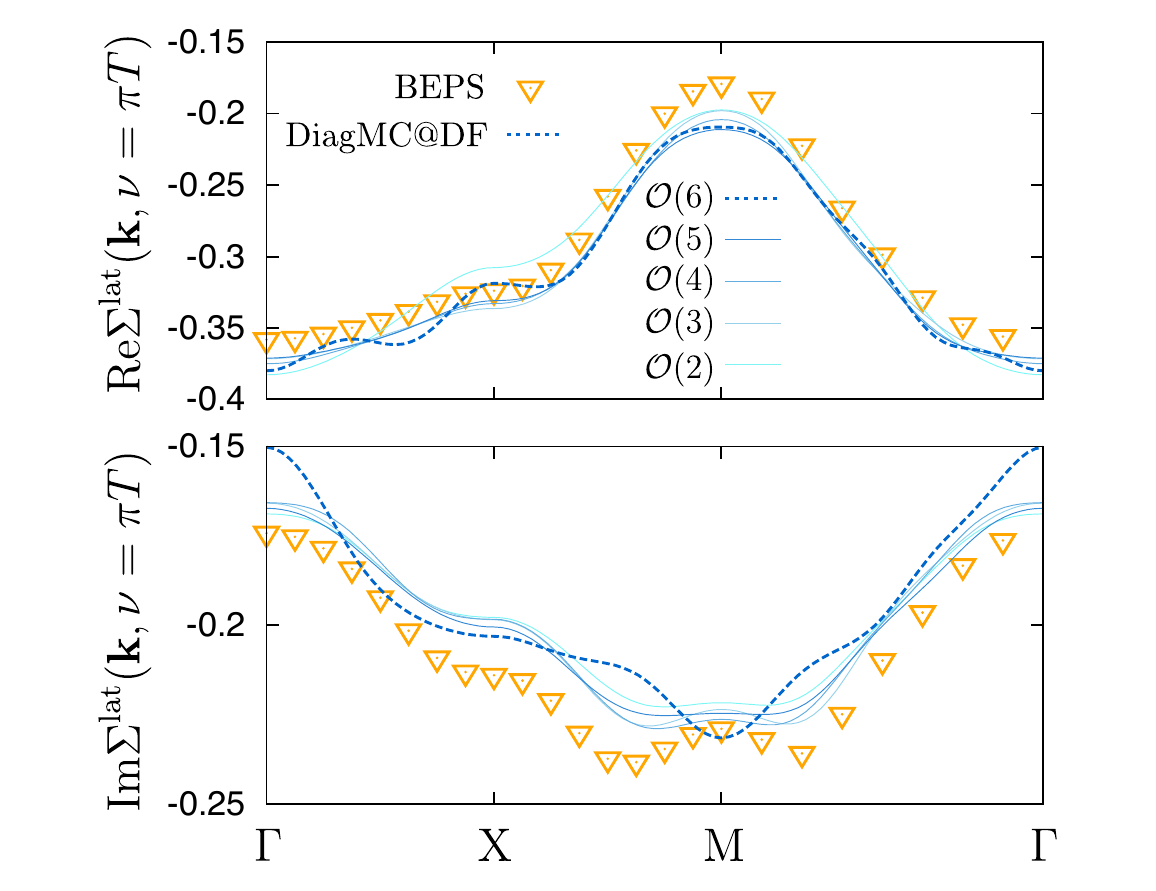}
    \end{center}
    \vspace{-0.5cm}
    \caption{\label{fig:u4b5_doped_path}
     $U/t=4, T/t=0.2, n\approx0.76$. Labels as in Fig.~\ref{fig:u4b2_doped_path},
    various expansion orders of DiagMC@DF are shown in blue.
    }
\end{figure}

We depart from half-filling and show in Figs.~\ref{fig:u4b2_doped_path} and~\ref{fig:u4b5_doped_path}
two benchmarks of BEPS against DiagMC@DF at interaction $U/t=4$ and temperatures $T/t=0.5$ and $0.2$, respectively.
The density is set to $n\approx0.76$.
The Matsubara cutoff corresponds to the half-filled case (see Sec.~\ref{sec:benchmark}).
In both calculations the lattice size is $16\times16$ sites,
results are shown for $N_\ell=5$ form factors.
Differences to calculations using $N_\ell=1$ or $N_\ell=9$ form factors are indiscernible (not shown),
underlining once again the rapid convergence of BEPS with the form factors and
the short-ranged property of the residual vertex $\Phi^\firr$, also away from half-filling.
At $T/t=0.5$ the agreement of BEPS and DiagMC@DF is excellent.
It is also reasonable at the lower temperature $T/t=0.2$,
but the statistical error of DiagMC@DF, its variation with the perturbation order,
and a difference in the densities~\footnote{
In the doped case we fixed the filling of the DMFT calculation to DiagMC@DF, $n=0.76$.
As a result, the density of the BEPS calculations is slightly off by $\pm0.008$
and it would be desirable to fix the chemical potential.
This requires to couple our BEPS implementation self-consistently to the {\sl w2dynamics }impurity solver,
which we leave for future work.} preclude a statement about the accuracy of BEPS for these parameters.

Finally, we note a peculiarity of the BEPS method that may somewhat impede its practicality in the short-term.
The method requires the {\sl complete} two-, three-, and four-point information of the AIM~\eqref{eq:aim},
including the particle-particle three-leg vertex $\lambda^\sing$
of the singlet channel and the corresponding susceptibility $\chi^\sing$,
defined in Eqs.~\eqref{eq:lambdasing} and~\eqref{eq:chi_sing}, respectively.
At half-filling we use a segment code~\cite{ALPS2,Hafermann12}
and obtain the particle-particle quantities from the charge channel via
particle-hole symmetry~\cite{Krien19-2}, however,
the doped case requires their measurement in a suitable CTQMC implementation.
We are unaware of a segment code~\cite{Gull11} that could handle the pair operator $c_\up c_\dn$
and we instead rely on the worm-sampling of the
W2DYNAMICS solver~\cite{Gunacker16,Tagliavini18,Wallerberger19},
which has however a larger statistical error than the segment code.
A better treatment of the particle-particle quantities is desirable, for example,
by using improved estimators~\cite{Kaufmann19},
exact diagonalization~\cite{Toschi07,Tanaka18,Tagliavini18},
or the numerical renormalization group (NRG,~\cite{Zitko09-2,Yang20}).

\section{Conclusions}\label{sec:conclusions}
We have introduced and applied to the two-dimensional Hubbard model a method for the summation of parquet diagrams for dual fermions~\cite{Rubtsov08} which substantially reduces the computational cost and increases the feasibility with respect to previous approaches.
The method makes use of the fact that the partial bosonization~\cite{Denz19} of the dual vertex function,
formalized in terms of the recently introduced single-boson exchange (SBE) decomposition~\cite{Krien19-2},
can be combined in a fruitful way with the traditional parquet formalism~\cite{Dominicis64,Dominicis64-2}.
Namely, as shown in the Appendices of this manuscript,
the parquet approximation for dual fermions can be cast \text{exactly} into
a set of parquet expressions for the residual vertex defined after the SBE decomposition explicitly treats single-boson-exchange diagrams.

This is a {significant} improvement because of useful properties of the residual vertex. 
In particular, it decays fast both in terms of Matsubara frequencies
{(\cite{Rohringer12,Wentzell2020,Krien19-2}, cf. Appendix~\ref{app:nu_convergence})}
and in terms of distances in the real space.
The latter property invites a truncated unity approximation~\cite{Eckhardt20}
at the level of the residual vertex,
whereas the full momentum-dependence of the single-boson exchange is retained.
As a result, we find across different {parameter} regimes that the electronic self-energy converges
rapidly with the number of form factors taken into account,
and significantly faster than in the TUPS method introduced in Ref.~\cite{Eckhardt20}.
{As for the frequency domain, we follow a similar philosophy of Ref.~\cite{Astretsov19} by evaluating the corresponding parquet expressions only for a small number of Matsubara frequencies. Our approach, however, preserves the essential spectral information of the underlying physical systems.}

In this work we have mostly focused on the description of the approach and to a %relatively light
{preliminary} application for the Hubbard model on up to $16\times16$ lattice sites.
{In fact,} we can currently reach a $32\times32$ lattice~\cite{Krien20-3} and % the
numerical aspects of the implementation can be {further} improved.

To {highlight} that our method corresponds to a merger of the SBE decomposition with TUPS,
we {coin it} boson exchange parquet solver (BEPS).
{As it has been recently shown~\cite{Krien20-2}, the versatility of the BEPS formalism allows for its application also to parquet-based approaches, such as the parquet approximation~\cite{Bickers04}, the D$\Gamma$A~\cite{Toschi07} and/or the QUADRILEX~\cite{Ayral16} formulated in terms of the {\sl original} (i.e., non dual) fermionic variables.
In this way, most of the numerical advantages described in this paper become available to all the abovementioned schemes. 
In this respect, let us note that while the physical content of a given parquet-based approach will not be excessively affected by the choice of formulating it in terms of the original or of the dual degrees of freedom, the latter procedure offers specific technical advantages, especially in the intermediate-to-strong-coupling regime.}

{In fact, the dual-fermion formulation of parquet-based schemes allows -per construction- to fully bypass the multiple divergences of the (local)
two-particle irreducible vertex, whose occurrence is rather ubiquitous in the phase diagrams of many-electron problems ~\cite{Schaefer13,Janis14,Gunnarsson16,Schaefer16,Ribic16,Gunnarsson17,Chalupa18,Thunstroem18,Springer20}.
As a result, the corresponding parquet decomposition~\cite{Gunnarsson16,Kauch19-3} of the electronic self-energy as well as of physical response functions remains well-behaved
at strong coupling, alleviating convergence problems of the parquet solver in regimes relevant for the experiment~\cite{Astretsov19}.}

We also notice how the developments that we proposed are intertwined
with the functional renormalization group methods (fRG). 
Two techniques often employed in the fRG framework, namely
the partial bosonization~\cite{Denz19,Bonetti20} and the truncated unity
approximation~\cite{Thomale13,Lichtenstein17,Eckhardt20,Hille20,Hille20-2}
are indeed instrumental to construct BEPS.
Vice versa, elements of our method could be useful for the fRG,
in particular, we find it plausible that a multi-loop fRG for dual fermions
in combination with partial bosonization could be cast into a
calculation scheme with properties similar to BEPS.
Indeed, for lattice fermions the multi-loop fRG
corresponds exactly to the summation of the parquet diagrams~\cite{Kugler18,Kugler18-2,Tagliavini19,Chalupa21}.
The groundwork for a combination of the fRG with strong-coupling theories like DMFT or dual fermions
was laid in Refs.~\cite{Taranto14,Wentzell15,Katanin19}.

{The methodological advancement provided by BEPS appears promising for extending the applicability of state-of-the-art parquet and fRG schemes to the most interesting regime of intermediate-to-strong local and nonlocal correlations.
In particular, we note that the BEPS implementations might considerably improve our non-perturbative description of the interplay between competing fluctuations, such as those originating from commensurate as well as  incommensurate magnetic and charge instabilities, or diverse pairing instabilities. In fact, while some of the these transitions have been investigated in the past within the ladder approximation~\cite{Otsuki14,Schaefer17,DelRe19,Kitatani20}, only a parquet treatment with sufficient momentum resolution might yield an equal-footing description of all competing fluctuations at play. Rather straightforward generalization of the procedure should also allow for the description of more complex magnetic instabilities, such as those towards a spin-spiral order.
On a longer-term perspective, BEPS might also provide a favorable framework to include non-local correlations on top of DMFT in magnetically/excitonic ordered phases~\cite{Sangiovanni06,Taranto12,Kunes19,Kunes20,Bonetti20-2}, as well as to treat multi-orbital systems~\cite{Toschi11} beyond the ladder approximation~\cite{Galler16,Kaufmann20,DelRe20}.

At the same time, it is questionable whether parquet resummations of nonlocal correlations can at all capture the resonating valence bond state or the spin-liquid phase.
These applications may require a cluster extension of BEPS to recover the short-ranged singlet physics nonperturbatively, in the spirit of the so-called multiscale approaches~\cite{Slezak09}.
Further, we expect that the convergence of the truncated-unity approximation applied
to the residual vertex $\Phi^\text{Uirr}$ may be slowed down when
this vertex develops a strong momentum dependence.
In the applications of the BEPS method it is therefore still important
to carefully verify the convergence of key observables with the number of form factors.
}

\acknowledgments
F.K. thanks A. Amaricci, S. Andergassen, C. Eckhardt, K. Held, C. Hille,
S. Huber, A. Katanin, A. Kauch, J. Kokalj, E.G.C.P. van Loon,
J. Mravlje, P. Prelov\v{s}ek, G. Rohringer, T. Sch\"afer,
M. Wallerberger and R. \v{Z}itko for encouraging discussions.
F.K. acknowledges financial support from the Slovenian Research Agency under project number N1-0088.
The present research was supported by the Austrian Science Fund (FWF) through project P32044.
P.C. and A.T. acknowledge financial support from the Austrian Science Fund (FWF) project number No. I 2794-N35.
A.V. acknowledges financial support from the Austrian Science Fund (FWF) project number No.~P31631.
M.C. acknowledges financial support from Ministero dell'Istruzione, dell'Universit\`a e della Ricerca under PRIN 2017 ”CEnTraL” and H2020 Framework Programme, under ERC Advanced Grant No. 692670 FIRSTORM.

\appendix

\section{SBE decomposition for dual fermions}\label{app:susc}
We explain how the SBE decomposition derived in Ref.~\cite{Krien19-2} can be formulated for dual fermions.
\subsection{Irreducible generalized susceptibility}\label{app:gsuscirr}
First, we define a dual generalized susceptibility as 
\begin{align}
{X}^\alpha_{kk'q}={X}^0_{kk'q}+\sum_{k_1k_2}{X}^0_{kk_1q}F^\alpha_{k_1k_2q}{X}^0_{k_2k'q}, \label{eq:gensusc}
\end{align}
where $F$ is the full vertex and ${X}^0_{kk'}=N\beta G_kG_{k+q}\delta_{kk'}$ is the bubble of dual fermions, respectively.
We denote as $\tilde{\Gamma}^{ph}$ the two-particle self-energy, i.e., the vertex which is irreducible with respect to horizontal particle-hole pairs. The generalized susceptibility satisfies the ladder equation 
\begin{align}
\hat{X}=\hat{X}^0+\hat{X}^0\hat{\tilde{\Gamma}}^{ph}\hat{X},
\end{align}
where we adopted a matrix notation with respect to the indices $k,k'$. Labels $\alpha, q$ are dropped.

The goal is to separate from $X$ and $F$ the diagrams that are reducible with respect to the
Hubbard interaction $U$, where we begin with the horizontal particle-hole channel.
For lattice fermions these reducible contributions arise from the leading term $U$
of the two-particle self-energy~\cite{Krien19}.
However, the dual two-particle self-energy $\tilde{\Gamma}^{ph}$ has many more $U$-reducible contributions,
since its leading term is the full vertex $f$ of the AIM~\eqref{eq:aim}.
The $U$-reducible contributions $\nabla^{ph}$ of the horizontal particle-hole channel can be separated off,
\begin{align}
    f^\alpha_{\nu\nu'\omega}=t^{ph,\alpha}_{\nu\nu'\omega}+\nabla^{ph,\alpha}_{\nu\nu'\omega}.
\end{align}
Hence, we subtract the $U$-reducible diagrams from the two-particle self-energy,
$S_{kk'q}^{ph}=\tilde{\Gamma}^{ph}_{kk'q}-\nabla^{ph}_{\nu\nu'\omega}$,
and define the following $\nabla$-\textit{irreducible} generalized susceptibility,
\begin{align}
    \hat{\Pi}=&\hat{X}^0+\hat{X}^0\hat{S}^{ph}\hat{\Pi}.\label{app:piladder}
\end{align}
The reducible and irreducible generalized susceptibilities are related as follows,
\begin{align}
\hat{X}=&\hat{\Pi}+\hat{\Pi}\,\hat{\nabla}^{ph}\hat{X},\notag\\
    \Leftrightarrow X_{kk'q}=&\Pi_{kk'q}+\sum_{k_1 k_2}\Pi_{kk_1q}\nabla^{ph}_{\nu_1\nu_2\omega}{X}_{k_2 k'q},
\end{align}
where the summation over matrix elements was made explicit in the second line.
We can now make use of the fact that $\nabla^{ph}$ depends on $\nu$ and $\nu'$ \textit{separately}, $\nabla^{ph}_{\nu\nu'\omega}={\lambda}^{\alpha}_{\nu\omega}w^\alpha_\omega\lambda^{\alpha}_{\nu'\omega}$,
where $\lambda$ is defined in Eq.~\eqref{eq:hedinvertex},
\begin{align}
    X_{kk'q}=&\Pi_{kk'q}+\left(\sum_{k_1}\Pi_{kk_1q}{\lambda}_{\nu_1\omega}\right)w_\omega\left(\sum_{k_2}\lambda_{\nu_2\omega}{X}_{k_2 k'q}\right).\label{eq:xfromxi}
\end{align}
This relation shows that if we take a trace $\sum_k$ over two-particle correlation functions for dual fermions,
it is natural to attach the impurity Hedin vertex $\lambda$ first. We do this when we take the trace over $k,k'$ in Eq.~\eqref{eq:xfromxi},
\begin{align}
    \frac{1}{2}X_q\equiv&\sum_{kk'}\lambda_{\nu\omega}X_{kk'q}{\lambda}_{\nu'\omega}
    =\sum_{kk'}\lambda_{\nu\omega}\Pi_{kk'q}{\lambda}_{\nu'\omega}\\
    +&\left(\sum_{kk_1}\lambda_{\nu\omega}\Pi_{kk_1q}{\lambda}_{\nu_1\omega}\right)w_\omega\left(\sum_{k_2k'}\lambda_{\nu_2\omega}{X}_{k_2 k'q}{\lambda}_{\nu'\omega}\right).\notag
\end{align}
We further define,
\begin{align}
\Pi_{q}\equiv\sum_{kk'}\lambda_{\nu\omega}\Pi_{kk'q}{\lambda}_{\nu'\omega},\label{app:pit}
\end{align}
and hence arrive at the \textit{algebraic} relation,
\begin{align}
X^\alpha_q=\frac{2\Pi^\alpha_q}{1-w^\alpha_\omega\Pi^\alpha_q}.\label{eq:xgeo}
\end{align}
The quantities $X$ and $\Pi$ naturally define the susceptibility and polarization of the dual fermions.

\subsection{SBE vertex}\label{app:sbevertex}
Now we separate the $U$-reducible contributions from the full vertex $F$.
To this end, we define a vertex part for the irreducible generalized susceptibility, similar to Eq.~\eqref{eq:gensusc},
\begin{align}
{\Pi}^\alpha_{kk'q}={X}^0_{kk'q}+\sum_{k_1k_2}{X}^0_{kk_1q}T^{ph,\alpha}_{k_1k_2q}{X}^0_{k_2k'q}. \label{eq:gensusc_pi}
\end{align}
We insert this relation and Eq.~\eqref{eq:gensusc} into Eq.~\eqref{eq:xfromxi} and cancel all bubbles $X^0$,
leading to the relation,
\begin{align}
    F^\alpha_{kk'q}=T^{ph,\alpha}_{kk'q}+{\Lambda}^{\alpha}_{kq}w^\alpha_\omega\Lambda^{\alpha,\text{red}}_{k'q},
    \label{eq:firr}
\end{align}
where we defined the three-leg vertices ${\Lambda}$ and $\Lambda^\text{red}$ as,
\begin{align}
  X^0_{kq}{\Lambda}_{kq}=&\sum_{k_1}\Pi_{kk_1q}{\lambda}_{\nu_1\omega},\label{app:lambdai}\\
  \Lambda^{\text{red}}_{k'q}X^0_{k'q}=&\sum_{k_2}\lambda_{\nu_2\omega}{X}_{k_2 k'q}.\label{app:lambda}
\end{align}
We like to eliminate $\Lambda^\text{red}$ in favor of $\Lambda$ in Eq.~\eqref{eq:firr}
and from Eq.~\eqref{eq:xfromxi} it follows indeed that
$\Lambda^\text{red}_{kq}={\Lambda_{kq}}/(1-w_\omega\Pi_q)$, hence
\begin{align}
    F^\alpha_{kk'q}=T^{ph,\alpha}_{kk'q}+{\Lambda}^{\alpha}_{kq}W^\alpha_q\Lambda^{\alpha}_{k'q},\label{eq:lwl}
\end{align}
where we defined the dual screened interaction as,
\begin{align}
   W^\alpha_q = \frac{w^\alpha_\omega}{1-w^\alpha_\omega\Pi^\alpha_q}.\label{app:wdual}
\end{align}

Finally, in Eq.~\eqref{eq:lwl} we identify the SBE vertex of the horizontal particle-hole channel,
i.e., Eq.~\eqref{eq:nabla_simple},
\begin{align}
\Delta^{ph,\alpha}_{kk'q}\equiv{\Lambda}^{\alpha}_{kq}W^\alpha_q\Lambda^{\alpha}_{k'q}.\label{app:nabla}
\end{align}
Combining Eqs.~\eqref{app:lambdai} and~\eqref{eq:gensusc_pi} leads to
\begin{align}
  \Lambda^{\alpha}_{kq} = \lambda^\alpha_{\nu\omega}+\sum_{k'}T^{ph,\alpha}_{kk'q}G_{k'}G_{k'+q}
  \lambda^\alpha_{\nu'\omega},\label{app:hedin_chsp}
\end{align}
and using Eqs.~\eqref{eq:lwl} and~\eqref{app:nabla} we arrive at Eq.~\eqref{eq:hedin_chsp} in the main text.

The remaining task is to find the vertices $\Delta^{\overline{ph}}$ and $\Delta^{pp}$
of the vertical particle-hole and particle-particle channels, respectively.
The first follows from the crossing relation in Eq.~\eqref{eq:cs}, 
the derivation of the latter proceeds along similar steps as in the Appendix of Ref.~\cite{Krien19-2},
leading to Eq.~\eqref{eq:nabla_simple_pp} [and Eq.~\eqref{eq:nabla_singlet} in particle-hole notation].
Removing $\Delta^{ph}$, $\Delta^{\overline{ph}}$, and $\Delta^{pp}$ from the full dual vertex function $F$,
and taking care of their double counting of the bare interaction~\cite{Krien19-2},
we call the remainder $\Phi^\firr$ and arrive at the SBE decomposition in Eq.~\eqref{eq:jib_simple}.

\section{Relation to parquet formalism}\label{app:relationtoparquet}
We relate the SBE decomposition to the parquet formalism.
The traditional parquet equation for the full vertex reads in particle-hole notation,
\begin{align}
    F^\alpha_{kk'q}=&\tilde{\Lambda}^{\text{firr},\alpha}_{kk'q}+\tilde{\Phi}^{ph,\alpha}_{kk'q}\label{app:parquet}\\
    -&\frac{1}{2}\tilde{\Phi}^{ph,\ch}_{k,k+q,k'-k}
    -\frac{3-4\delta_{\alpha,\sz}}{2}\tilde{\Phi}^{ph,\sz}_{k,k+q,k'-k}\notag\\
    +&\frac{1-2\delta_{\alpha,\sz}}{2}\tilde{\Phi}^{pp,\sing}_{kk',k+k'+q}
    +\frac{3-2\delta_{\alpha,\sz}}{2}\tilde{\Phi}^{pp,\trip}_{kk',k+k'+q}.\notag
\end{align}
Here, $\tilde{\Lambda}^{\text{firr}}$ is the fully irreducible vertex in the sense of the traditional parquet formalism~\cite{Rohringer12}, which implies it is irreducible with respect to insertions of particle-hole and particle-particle pairs.
The vertices $\tilde{\Phi}^{ph(pp)}$ are \textit{reducible} in this sense
(either in a particle-hole or particle-particle channel).
All quantities which are reducible or irreducible in the sense of the traditional
parquet formalism are marked with a tilde.
\textit{In particular, }$\tilde{\Lambda}^{\text{firr}}, \tilde{\Phi}$ \textit{should not be confused with the vertex} $\Phi^\firr$, \textit{which is (fully) irreducible with respect to the bare interaction} $U$~\cite{Krien19-2,Krien19-3}.
A closed set of equations is obtained in combination with the Bethe-Salpeter equations,
\begin{align}
  F^\alpha_{kk'q}=&\tilde{\Gamma}^{ph,\alpha}_{kk'q}+\tilde{\Phi}^{ph,\alpha}_{kk'q}, \;\;\alpha=\ch,\sz,\label{app:bseph}\\
  F^\delta_{kk'q}=&\tilde{\Gamma}^{pp,\delta}_{kk'q}+\tilde{\Phi}^{pp,\delta}_{kk'q}, \;\;\delta=\sing,\trip.\label{app:bsepp}
\end{align}
where $\tilde{\Gamma}$ is irreducible with respect to particle-hole or particle-particle pairs.
In the SBE decomposition the vertex is split according to Eq.~\eqref{eq:lwl},
\begin{align}
    F^\alpha_{kk'q}=&T^{ph,\alpha}_{kk'q}+\Delta^{ph,\alpha}_{kk'q},\label{app:lwlph}\\
    F^\delta_{kk'q}=&T^{pp,\delta}_{kk'q}+\Delta^{pp,\delta}_{kk'q}.\label{app:lwlpp}
\end{align}
The vertices $T$ are irreducible with respect to the bare interaction in a particular channel (and therefore $\Delta^{pp,t}=0$ for the triplet channel),
they obey the following Bethe-Salpeter-like equations [cf. Eq.~\eqref{app:piladder}],
\begin{align}
    T^{ph,\alpha}_{kk'q}=&S^{ph,\alpha}_{kk'q}+M^{ph,\alpha}_{kk'q},\label{app:fimph}\\
    T^{pp,\delta}_{kk'q}=&S^{pp,\delta}_{kk'q}+M^{pp,\delta}_{kk'q},\label{app:fimpp}
\end{align}
where the vertices $S$ and $M$ are defined as~\footnote{
Regarding prefactor $\mp\frac{1}{2}$ of $pp$-channel see Ref.~\cite{Rohringer12}.},
\begin{align}
    S^{ph,\alpha}_{kk'q}=&\tilde{\Gamma}^{ph,\alpha}_{kk'q}-\nabla^{ph,\alpha}_{\nu\nu'\omega},\label{app:sph}\\
    S^{pp,\delta}_{kk'q}=&\tilde{\Gamma}^{pp,\delta}_{kk'q}-\nabla^{pp,\delta}_{\nu\nu'\omega},\label{app:spp}\\
    M^{ph,\alpha}_{kk'q}=&\sum_{k''}S^{ph,\alpha}_{kk''q}G_{k''}G_{k''+q}T^{ph,\alpha}_{k''k'q},\\
    M^{pp,\delta}_{kk'q}=&\mp\frac{1}{2}\sum_{k''}S^{pp,\delta}_{kk''q}G_{k''}G_{q-k''}T^{pp,\delta}_{k''k'q}.
\end{align}
We now express the reducible vertices $\tilde{\Phi}$ of the traditional parquet formalism in terms of the new vertices $M$.
Combining the previous equations we arrive at,
\begin{align}
    \tilde{\Phi}^{ph,\alpha}_{kk'q}=&\Delta^{ph,\alpha}_{kk'q}-\nabla^{ph,\alpha}_{\nu\nu'\omega}+M^{ph,\alpha}_{kk'q},\label{app:phimph}\\
    \tilde{\Phi}^{pp,\delta}_{kk'q}=&\Delta^{pp,\delta}_{kk'q}-\nabla^{pp,\delta}_{\nu\nu'\omega}+M^{pp,\delta}_{kk'q},\label{app:phimpp}
\end{align}
which leads to $M^{pp,\trip}=\tilde{\Phi}^{pp,\trip}$ for the triplet channel.

\section{Parquet approximation}\label{app:mbe}
We reformulate the parquet approximation for dual fermions in terms of
parquet expressions for the residual vertex $\Phi^\firr$.
The parquet approximation for dual fermions corresponds to,
\begin{align}
    \tilde{\Lambda}^{\text{firr}}_{kk'q}\approx f_{\nu\nu'\omega},
\end{align}
that is, the fully irreducible vertex of the traditional parquet formalism
is given by the full vertex of the impurity model.
We insert this approximation and Eqs.~\eqref{app:phimph} and~\eqref{app:phimpp} into the parquet equation~\eqref{app:parquet} and compare with the SBE decomposition~\eqref{eq:jib_simple},
\begin{align}
    F^\alpha_{kk'q}=&{\Phi}^{\firr,\alpha}_{kk'q}+\Delta^{ph,\alpha}_{kk'q}\label{app:jib_full}\\
    -&\frac{1}{2}\Delta^{ph,\ch}_{k,k+q,k'-k}
    -\frac{3-4\delta_{\alpha,\sz}}{2}\Delta^{ph,\sz}_{k,k+q,k'-k}\notag\\
    +&\frac{1-2\delta_{\alpha,\sz}}{2}\Delta^{pp,\sing}_{kk',k+k'+q}-2U^\alpha=\text{Eq.~\eqref{app:parquet}}.\notag
\end{align}
Using also the corresponding SBE decomposition for the impurity vertex $f$ in Eq.~\eqref{eq:jib_aim} all vertices
$\nabla, \Delta$ and the bare interaction $U$ cancel out, and we arrive at the following parquet expression,
\begin{align}
    \Phi^{\firr,\alpha}_{kk'q}=&\varphi^{\firr,\alpha}_{\nu\nu'\omega}+M^{ph,\alpha}_{kk'q}\label{app:parquet_firr}\\
    -&\frac{1}{2}M^{ph,\ch}_{k,k+q,k'-k}
    -\frac{3-4\delta_{\alpha,\sz}}{2}M^{ph,\sz}_{k,k+q,k'-k}\notag\\
    +&\frac{1-2\delta_{\alpha,\sz}}{2}M^{pp,\sing}_{kk',k+k'+q}
    +\frac{3-2\delta_{\alpha,\sz}}{2}M^{pp,\trip}_{kk',k+k'+q}.\notag
\end{align}
This corresponds to an exact reformulation of the parquet approximation for dual fermions.
For the evaluation we further need the ladder kernel $S$.
We use Eqs.~\eqref{app:lwlph}-\eqref{app:fimpp} and Eq.~\eqref{app:jib_full}
to express the particle-hole kernel $S^{ph}$ in Eq.~\eqref{app:sph} as,
\begin{align}
    S^{ph,\alpha}_{kk'q}=&\Phi^{\firr,\alpha}_{kk'q}-M^{ph,\alpha}_{kk'q}\notag\\
    -&\frac{1}{2}\Delta^{ph,\ch}_{k,k+q,k'-k}
    -\frac{3-4\delta_{\alpha,\sz}}{2}\Delta^{ph,\sz}_{k,k+q,k'-k}\notag\\
    +&\frac{1-2\delta_{\alpha,\sz}}{2}\Delta^{pp,\sing}_{kk',k+k'+q}-2U^\alpha\label{app:sexact}.
\end{align}
Similar steps lead to Eqs.~\eqref{eq:pp_kernel_sing} and~\eqref{eq:pp_kernel_trip} for the particle-particle channels.

\section{Ladder kernel in form-factor basis}\label{app:formf}
We show in an exemplary way the calculation of the different components of the ladder kernel in Eqs.~\eqref{eq:ph_kernel_ch}-\eqref{eq:pp_kernel_trip} in
the form-factor basis. For the particle-hole kernel (we drop frequency and flavor labels),
\begin{align}
    S^{ph}(\lv_1,\lv_2,\qv)=&\sum_{\kv\kv'}\psi(\lv_1,\kv)S^{ph}(\kv,\kv',\qv)\psi(\lv_2,\kv').\label{app:sellell}
\end{align}
We use the truncated unity to avoid the full momentum-dependence of four-point vertices.
For example, following Ref.~\cite{Eckhardt20}, the contribution of the MBE vertex $M^{ph}$ on the right-hand-side
of Eq.~\eqref{eq:ph_kernel_ch} can be brought into the form,
\begin{widetext}
\begin{align}
     &\sum_{\kv\kv'}\psi(\lv_1,\kv)M^{ph}(\kv,\kv+\qv,\kv'-\kv)\psi(\lv_2,\kv')\notag\\
    =&\sum_{\kv\qv'}\psi(\lv_1,\kv)M^{ph}(\mathcal{S}[\kv],\mathcal{S}[\kv+\qv],\mathcal{S}[\qv'])\psi(\lv_2,\kv+\qv')\notag\\
    =&\sum_{\kv\qv'}\sum_{\lv_3\lv_4}\psi(\lv_1,\kv)\psi(\lv_3,\mathcal{S}[\kv])
    M^{ph}(\lv_3,\lv_4,\mathcal{S}[\qv'])\psi(\lv_4,\mathcal{S}[\kv+\qv])\psi(\lv_2,\kv+\qv')\notag\\
    =&\sum_{\qv'}\sum_{\lv_3\lv_4}\Psi_\mathcal{S}(\lv_1,\lv_2,\lv_3,\lv_4,\qv',\qv)M^{ph}(\lv_3,\lv_4,\mathcal{S}[\qv'])\label{app:tups_psi}
\end{align}
\end{widetext}
From the first to the second line we performed a shift $\qv'=\kv'-\kv$ and introduced a symmetry operation
$\mathcal{S}$ of the point-group, which is chosen to project the momentum $\qv'$ into the irreducible Brillouin zone
(the same operation therefore needs to be applied to the other
two momentum arguments of $M^{ph}$, see Ref.~\cite{Thomale13}).
In the third line $M^{ph}$ was transformed into the form-factor basis.
In the fourth line the four form factors were collected in the quantity,
\begin{widetext}
\begin{align}
\Psi_\mathcal{S}(\lv_1,\lv_2,\lv_3,\lv_4,\qv',\qv)
=&\sum_\kv\psi(\lv_1,\kv)\psi(\lv_2,\kv+\qv')\psi(\lv_3,S[\kv])\psi(\lv_4,S[\kv+\qv]).
\end{align}
\end{widetext}
In practice this quantity is calculated once at the beginning of the calculation,
keeping only a number $N_\ell$ of form factors.
The symmetry operation $\mathcal{S}$ allows to perform the summation over $\qv'$ in Eq.~\eqref{app:tups_psi}
only over the irreducible Brillouin zone rather than the full one.
The other components $M$ of the ladder kernel $S$ are handled analogously.
In this way, we avoid the storage of the MBE vertex $M(\kv,\kv',\qv)$ of size $N_k^2N_q^\text{irr}$
and store only $M(\lv,\lv',\qv)$ which has the size $N_\ell^2N_q^\text{irr}$, where $N_q^\text{irr}\approx N_q/8$
is the size of the irreducible Brillouin zone, see Ref.~\cite{Thomale13} for further information.
On the other hand, the full momentum-dependence of the SBE vertex $\Delta(\kv,\kv',\qv)$ can be stored efficiently,
since it is parameterized by the fermion-boson coupling $\Lambda(\kv,\qv)$
and the screened interaction $W(\qv)$ [cf. Sec.~\ref{sec:strategy}].
Hence, at each iteration we calculate the contribution of $\Delta$ to $S$
explicitly, performing the $\kv,\kv'$ summations in Eq.~\eqref{app:sellell}~\footnote{
In principle, we could avoid the transformation of $\Delta$ in Eq.~\eqref{app:sellell} and instead transform
$\Lambda(\kv,\qv)$ to the form-factor basis with respect to $\kv$,
obtain $\Delta^{ph}(\lv,\lv',\qv)=\Lambda(\lv,\qv)W(\qv)\Lambda(\lv',\qv)$,
and then follow the steps in Eq.~\eqref{app:tups_psi} to treat the momentum shifts of $\Delta$
in the ladder kernel $S$ given by Eqs.~\eqref{eq:ph_kernel_ch}-\eqref{eq:pp_kernel_trip}.
One may thus avoid $\kv$-dependence of $\Lambda$ alltogether,
but the performance gain is limited and the procedure introduces a further truncation error
which worsens the convergence with respect to the form-factor cutoff $N_\ell$.
We therefore use the truncated unity only to treat the momentum shifts of the vertex $M$.}.

\section{Frequency convergence}\label{app:nu_convergence}
\begin{figure}
    \begin{center}
         \includegraphics[width=0.45\textwidth]{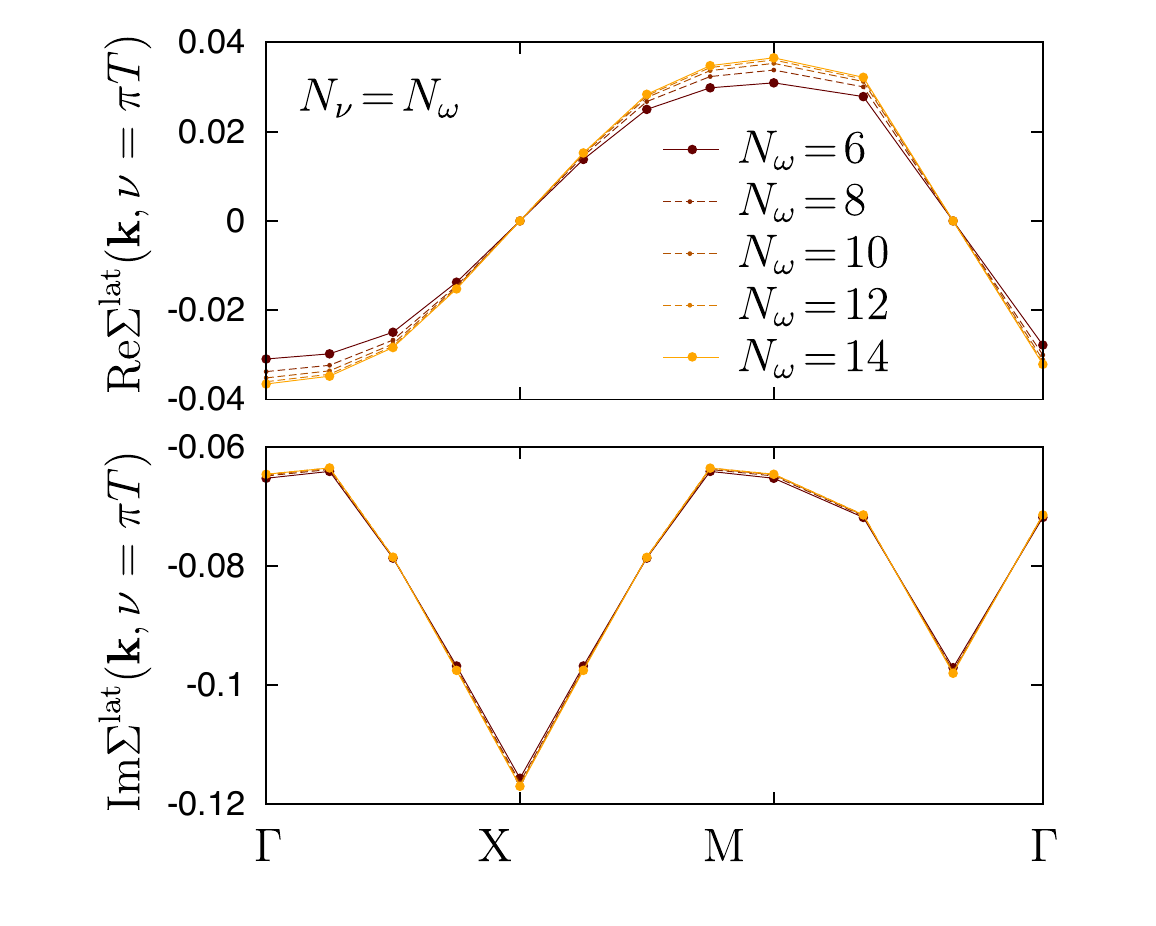}
    \end{center}
    \vspace{-0.5cm}
    \caption{\label{fig:u2b5_path_Nw}
     Self-energy at half-filling, $U/t=2, T/t=0.2$ for different values of the Matsubara cutoff.}
\end{figure}

Fig.~\ref{fig:u2b5_path_Nw} shows the convergence of the BEPS self-energy
with the Matsubara cutoff $N_\nu=N_\omega$ for a calculation at half-filling, $U/t=2, T/t=0.2$.
The lattice size is set to only $8\times8$,
leading to a finite-size effect, and we use only one form factor ,
which is however not relevant for the frequency convergence.
In the case that a Matsubara label exceeds the cutoff the respective quantity is
set to a default value, namely, $G\rightarrow G^0$, $\Sigma\rightarrow0$, $\Lambda\rightarrow1$,
$W^\alpha\rightarrow U^\alpha$, $\Phi^\firr\rightarrow0$, $M\rightarrow0$
(fermionic Matsubara indices of four-point vertices like $\Phi^\firr$ and $M$ run from $-N_\nu/2$ to $N_\nu/2-1$).
Only the quantities $\Sigma^\text{DMFT}$ and $g_\nu$ are defined on a larger Matsubara grid
($64$ frequencies in practice).

\bibliography{main}

\end{document}